\documentclass[12pt,prb,onecolumn]{revtex4}%
\usepackage{amsmath}
\usepackage{graphicx}%
\usepackage{amsfonts}%
\usepackage{amssymb}

\begin{document}

\title{Two-Dimensional Optical Spectroscopy of Excitons in Semiconductor Quantum
Wells: Liouville-space pathway analysis}
\author{Lijun Yang$^{\text{1}}$, Igor V. Schweigert$^{\text{1}}$, Steven T.
Cundiff$^{\text{2}}$, Shaul Mukamel$^{\text{1}}$}

\begin{abstract}
We demonstrate how dynamic correlations of heavy-hole and light-hole excitons
in semiconductor quantum wells may be investigated by two dimensional
correlation spectroscopy (2DCS). \ The coherent response to three femtosecond
optical pulses is predicted to yield cross (off-diagonal) peaks that contain
direct signatures of many-body two-exciton correlations. \ Signals generated
at various phase-matching directions are compared.

\end{abstract}
\maketitle

\address{$^{\text{1}}$\textit{Chemistry department, University of California ,
Irvine, CA, 92697-2025, United States}\\ $^{\text{2}}$\textit{JILA, University
of Colorado and National Institute of Standards and Technology, Boulder,
Colorado 80309-0440, United States }}

\section{Introduction}

Understanding the signatures of many-body interactions in the nonlinear
optical response of semiconductors is an important fundamental problem with
implications to all-optical and electro-optical device applications.\cite{Li}
\ The linear response to a weak optical field is well described by a model of
non-interacting quasiparticles. \ However, residual interactions, not
accounted for by these quasiparticles, can considerably affect the nonlinear
response. \ Similar to Frenkel excitons in molecular crystals and
aggregates,\cite{paper386} \ Coulomb correlations among quasiparticles can
dominate the nonlinear optical response of semiconductors, in marked contrast
to the behavior of atomic
systems.\cite{H-K,Chemla,AxtRMP,Wegener,Leo90,Lindberg92,Bonitz,Hohenster,Pereira,Ivanov,kwong}
\ 

The coherent ultrafast response and many-body correlations in semiconductor
heterostructures have been studied extensively in the past two decades.
\cite{AxtRMP,Shahbook,Haug,Rossi,Axt PRL,Bartels,Hader,Osterich
I,Banyai1,Sieh,Shapiro,Lyssenko,infinite-order} \ Due to\ various dephasing
and relaxation mechanisms, the coherent response usually persists only on the
tens of picosecond time scale. \ Optical spectra such as the linear
absorption, pump-probe and Four-wave mixing (FWM) are commonly displayed as a
function of a single (time or frequency) variable, and hence provide a
one-dimensional (1D) projection of the microscopic information. \ 1D spectra
are hard to interpret in systems with many congested energy levels.\ The
spectroscopic signatures of complex many-body dynamics projected on a 1D
spectral plot strongly overlap and may not be easily identified. \ For
example, when 1D techniques are employed in III-V semiconductor quantum wells
(QWs), it is difficult to pinpoint the signatures of two-excitons due to the
small (1 to 2 meV) two-exciton interaction energies, i.e., bound two-exciton
binding energy and unbound two-exciton scattering
energy.\cite{Smirl,Phillips,Meier2} \ Two-exciton effects can be easily masked
by homogeneous and inhomogeneous line broadening. \ When both light-hole
(LH)\ and heavy-hole (HH) excitons and their interactions are taken into
account, the situation becomes even more complicated since there are different
types of two-excitons such as LH, HH and mixed
two-excitons.\cite{Meier2,Wagner} \ Even in II-VI semiconductor QWs, where the
two-exciton interaction energies are much larger, ( from several to several
tens of meV),\cite{Wagner} it is still hard to resolve them. \ 

Multi-dimensional
spectroscopy\cite{paper386,tanimura1,chernyak96,chernyak,Zhang,paper487,WeiJcp06}
in which the optical response is recorded and displayed versus several
arguments, can overcome these limitations by separating the signatures of
different pathways of the density matrix, known as Liouville space
pathways.\cite{298} \ Because different pathways are usually connected with
specific couplings, one can resolve different coherences and many-body
interactions by focusing on different peaks in multi-dimensional correlation
plots. \ Optical and infrared 2D Correlation Spectroscopy (2DCS) is a
femtosecond analogue of multi-dimensional nuclear magnetic resonance
(NMR)\cite{217,218} that has been shown to be very powerful for probing the
structure and dynamics in chemical and biological
systems.\cite{ShaulPNAS1,WeiPNAS1,WeiJCP,DariousJCP,Fayer,Hamm,Fleming,Hochstrasser}
\ These techniques can reveal the coupling strength of elementary excitation
through \textit{cross-peaks}. \ After several
attempts\cite{KochM,Lyssenko1,Cundiff96PRL,Euteneuer,Finger} to go beyond 1D
techniques, the first experimental implementation of 2DCS to investigate the
many-body Coulomb interactions among LH\ and heavy-hole HH excitons in
semiconductors has been reported recently.\cite{Cundiff06PRL,CundiffCPL} \ 

In this work, we present the basic principles\cite{paper386,paper487} of 2DCS
in semiconductor QWs and show how it can be used to study the couplings among
LH and HH\ excitons in a GaAs single quantum well. \ The same methodology
previously applied to Frenkel excitons (hard-core bosons) in molecular
aggregates or anharmonic vibrations (soft-core bosons) in
proteins\cite{paper386,paper487} can be extended to Wannier
excitons.\cite{chernyak96,chernyak,Yokojima,Zhang1} \ In Section II, we survey
the possible third-order 2D techniques of excitons. \ The double-sided Feynman
diagrams and the corresponding sum-over-states expressions for 2DCS in III-V
semiconductor quantum wells are presented in Section III. \ These guide the
analysis of the numerical calculations presented in later sections. \ In
Section IV, we present the multiband Hamiltonian and the nonlinear exciton
equations used for the numerical calculation of the 2DCS signals. \ The
numerical calculations are analyzed in Section V. \ Finally, in Section VI, we
summarize our findings.

\section{Liouville-Space Pathway Analysis for 2D Correlation Spectroscopy of Excitons}

Impulsive 2DCS signals generated by well-separated femtosecond pulses may be
calculated using sum-over-states expressions which provide insights into the
origin of various peaks associated with different coherences and many-body
interactions. \ These expressions will be used in the following sections to
analyze the 2D signals obtained by numerical solution of equations of motion
with finite-envelope pulses.

We consider a three-pulse sequence where the electric field is given by \ %

\begin{equation}
\mathbf{E}(\mathbf{r},t)=\sum_{j=1}^{3}\left[  \mathbf{e}_{j}\mathcal{E}%
_{j}^{+}(t-\tau_{j})e^{i\left(  \mathbf{k}_{j}\cdot\mathbf{r}-\omega
_{j}t\right)  }+\mathbf{e}_{j}\mathcal{E}_{j}^{-}(t-\tau_{j})e^{-i\left(
\mathbf{k}_{j}\cdot\mathbf{r}-\omega_{j}t\right)  }\right]  .\label{efield}%
\end{equation}
Here $\mathcal{E}_{j}^{+}$ ($\mathcal{E}_{j}^{-}=\left(  \mathcal{E}_{j}%
^{+}\right)  ^{\ast}$) is the envelope of the positive(negative)-frequency
component of the $j$-th pulse centered at $\tau_{j}$, with carrier frequency
$\omega_{j}$, polarization unit vector $\mathbf{e}_{j}$, and wavevector
$\mathbf{k}_{j}$. \ The ability to vary the envelopes, polarization
directions, durations and time intervals, tune the frequencies and even
control the phases of optical pulses, provides a broad class of possible
techniques. \ By scanning these various parameters it is possible to design a
multiple-pulse technique \cite{paper386,paper487} for a specific application,
as is done in NMR.\cite{NMR} \ 

The total Hamiltonian is%
\[
H=H_{0}-\mathbf{\mu\cdot E}\left(  \mathbf{r},t\right)  ,
\]
where $H_{0}$ is the free system Hamiltonian and $\mathbf{\mu}$ is the dipole
operator. \ The third-order contribution to the system's polarization induced
by the interaction with the field can be written as
\begin{equation}
P_{e_{4}}(\mathbf{r},t)=\int_{0}^{\infty}\int_{0}^{\infty}\int_{0}^{\infty
}dt_{1}dt_{2}dt_{3}R_{e_{4}e_{3}e_{2}e_{1}}^{(3)}(t_{3},t_{2},t_{1})E_{e_{3}%
}(\mathbf{r},t-t_{3})E_{e_{2}}(\mathbf{r},t-t_{3}-t_{2})E_{e_{1}}%
(\mathbf{r},t-t_{3}-t_{2}-t_{1}),\label{polariz}%
\end{equation}
Here, $R^{(3)}$ is the third-order response function calculated using
time-dependent perturbation theory%
\begin{equation}
R_{e_{4}e_{3}e_{2}e_{1}}^{(3)}(t_{3},t_{2},t_{1})=i^{3}\left\langle \left[
\left[  \left[  \hat{\mathbf{\mu}}_{e_{4}}(t),\hat{\mathbf{\mu}}_{e_{3}%
}(t-t_{3})\right]  ,\hat{\mathbf{\mu}}_{e_{2}}(t-t_{3}-t_{2})\right]
,\hat{\mathbf{\mu}}_{e_{1}}(t-t_{3}-t_{2}-t_{1})\right]  \right\rangle
\end{equation}
where $\hat{\mathbf{\mu}_{e_{j}}}(t)$ are interaction picture operators.
\ $R^{(3)}$ is a tensor of rank four, but for clarity we hereafter suppress
the tensor indices $e_{j}$. \ In impulsive experiments where the field
envelopes are much shorter than the delay periods and the exciton dynamics
timescale, the system first interacts with the $\mathbf{k}_{1}$ pulse, then
with $\mathbf{k}_{2}$ and finally with $\mathbf{k}_{3}$, and the time
integrations in Eq. (\ref{polariz}) can be eliminated. \ We then get%

\begin{equation}
P^{(3)}(\mathbf{r},t)=R^{(3)}(t_{3},t_{2},t_{1})\mathcal{E}_{3}^{\pm
}\mathcal{E}_{2}^{\pm}\mathcal{E}_{1}^{\pm}e^{i(\pm\mathbf{k}_{1}\pm
\mathbf{k}_{2}\pm\mathbf{k}_{3})\mathbf{r}}e^{-i(\pm\omega_{1}\pm\omega_{2}%
\pm\omega_{3})t}e^{i(\pm\omega_{1}\pm\omega_{2}\pm\omega_{3})t_{3}}%
e^{i(\pm\omega_{1}\pm\omega_{2})t_{2}}e^{\pm i\omega_{1}t_{1}}.\nonumber
\end{equation}
We shall recast the polarization in the form%
\begin{equation}
P^{(3)}(\mathbf{r},t)=\sum_{s}^{4}P_{s}(t_{3},t_{2},t_{1})e^{i\mathbf{k}%
_{s}\mathbf{r}-i\omega_{s}t}+c.c.,
\end{equation}
where the signal wavevector $\mathbf{k}_{s}$ can assume one of the four
phase-matching values: $\ \mathbf{k}_{\mathbf{I}}=-\mathbf{k}_{1}%
+\mathbf{k}_{2}+\mathbf{k}_{3}$, $\mathbf{k}_{\mathbf{II}}=\mathbf{k}%
_{1}-\mathbf{k}_{2}+\mathbf{k}_{3}$, $\mathbf{k}_{\mathbf{III}}=\mathbf{k}%
_{1}+\mathbf{k}_{2}-\mathbf{k}_{3}$\ and $\mathbf{k}_{\mathbf{IV}}%
=\mathbf{k}_{1}+\mathbf{k}_{2}+\mathbf{k}_{3}$. \ 

By employing heterodyne detection and controlling the relative phases of the
heterodyne and incident fields, one can detect the signal field itself (both
amplitude and phase)\ rather than merely its intensity (modulus square). \ In
thin samples, propagation effects are negligible and the signal field is
proportional to the polarization field. \ Hereafter we denote the signal field
amplitudes generated along $\mathbf{k}_{j}$ as $\mathbf{S}_{j}$, where $j=$I,
II, III and IV. \ These are given by \
\begin{align}
S_{\mathbf{I}}\left(  t_{3},t_{2},t_{1}\right)   &  =iR_{\mathbf{I}}%
^{(3)}(t_{3},t_{2},t_{1})\mathcal{E}_{3}^{+}\mathcal{E}_{2}^{+}\mathcal{E}%
_{1}^{-}e^{i(-\omega_{1}+\omega_{2}+\omega_{3})t_{3}}e^{i(-\omega_{1}%
+\omega_{2})t_{2}}e^{-i\omega_{1}t_{1}},\label{1}\\
S_{\mathbf{II}}\left(  t_{3},t_{2},t_{1}\right)   &  =iR_{\mathbf{II}}%
^{(3)}(t_{3},t_{2},t_{1})\mathcal{E}_{3}^{+}\mathcal{E}_{2}^{-}\mathcal{E}%
_{1}^{+}e^{i(\omega_{1}-\omega_{2}+\omega_{3})t_{3}}e^{i(\omega_{1}-\omega
_{2})t_{2}}e^{i\omega_{1}t_{1}},\label{2}\\
S_{\mathbf{III}}\left(  t_{3},t_{2},t_{1}\right)   &  =iR_{\mathbf{III}}%
^{(3)}(t_{3},t_{2},t_{1})\mathcal{E}_{3}^{-}\mathcal{E}_{2}^{+}\mathcal{E}%
_{1}^{+}e^{i(\omega_{1}+\omega_{2}-\omega_{3})t_{3}}e^{i(\omega_{1}+\omega
_{2})t_{2}}e^{i\omega_{1}t_{1}},\label{3}\\
S_{\mathbf{IV}}\left(  t_{3},t_{2},t_{1}\right)   &  =iR_{\mathbf{IV}}%
^{(3)}(t_{3},t_{2},t_{1})\mathcal{E}_{3}^{+}\mathcal{E}_{2}^{+}\mathcal{E}%
_{1}^{+}e^{i(\omega_{1}+\omega_{2}+\omega_{3})t_{3}}e^{i(\omega_{1}+\omega
_{2})t_{2}}e^{i\omega_{1}t_{1}}.\label{4}%
\end{align}

The exciton level-scheme\cite{paper386,paper487} for the two-band model of
semiconductor shown in Fig. 1(top right ) consists of three manifolds: the
ground state ($g)$, single-exciton states ($e$) and two-exciton states ($f)$.
\ The dipole operator only connects the $g$\ to $e$\ and $e$\ to
$f$\ manifolds. \ Within the rotating wave approximation (RWA),
$R_{\mathbf{IV}}^{(3)}$ vanishes for this model, the response functions
$R_{j}^{(3)}$ [Eqs. (\ref{1}) to (\ref{3})] for the other techniques are
represented by the double-sided Feynman diagrams\cite{paper386,298} shown in
Fig. 1. \ 

The sum-over-states expression for $R_{\mathbf{I}}^{(3)}$ is%
\begin{align}
&  R_{\mathbf{I}}^{(3)}(t_{3},t_{2},t_{1})\label{specifRes}\\
&  =i^{3}\sum_{e,e^{\prime}}(\mathbf{e}_{1}\cdot\boldsymbol{\mu}_{ge^{\prime}%
})(\mathbf{e}_{2}\cdot\boldsymbol{\mu}_{e^{\prime}g})(\mathbf{e}_{4}%
\cdot\boldsymbol{\mu}_{ge})(\mathbf{e}_{3}\cdot\boldsymbol{\mu}_{eg}%
)e^{-(i\omega_{eg}+\Gamma_{eg})t_{3}-\Gamma_{gg}t_{2}+(i\omega_{e^{\prime}%
g}-\Gamma_{e^{\prime}g})t_{1}}\nonumber\\
&  +i^{3}\sum_{e,e^{\prime}}(\mathbf{e}_{1}\cdot\boldsymbol{\mu}_{ge^{\prime}%
})(\mathbf{e}_{3}\cdot\boldsymbol{\mu}_{e^{\prime}g})(\mathbf{e}_{4}%
\cdot\boldsymbol{\mu}_{ge})(\mathbf{e}_{2}\cdot\boldsymbol{\mu}_{eg}%
)e^{-(i\omega_{eg}+\Gamma_{eg})t_{3}-(i\omega_{ee^{\prime}}+\Gamma
_{ee^{\prime}})t_{2}+(i\omega_{e^{\prime}g}-\Gamma_{e^{\prime}g})t_{1}%
}\nonumber\\
&  -i^{3}\sum_{e,e^{\prime},f}(\mathbf{e}_{1}\cdot\boldsymbol{\mu}%
_{ge^{\prime}})(\mathbf{e}_{4}\cdot\boldsymbol{\mu}_{e^{\prime}f}%
)(\mathbf{e}_{3}\cdot\boldsymbol{\mu}_{fe})(\mathbf{e}_{2}\cdot\boldsymbol
{\mu}_{eg})e^{-(i\omega_{fe^{\prime}}+\Gamma_{fe^{\prime}})t_{3}%
-(i\omega_{ee^{\prime}}+\Gamma_{ee^{\prime}})t_{2}+(i\omega_{e^{\prime}%
g}-\Gamma_{e^{\prime}g})t_{1}},\nonumber
\end{align}
where $\omega_{vv^{\prime}}=\varepsilon_{v}-\varepsilon_{v^{\prime}}$
($\nu,\nu^{\prime}=g,e,e^{\prime},f$ ) is the frequency and $\Gamma_{\nu
\nu^{\prime}}$ are the dephasing rate of the $\nu\longrightarrow\nu^{\prime}$
transition. \ The three terms correspond respectively to diagrams (i), (ii)
and (iii). \ 

The 2D signal is displayed in the frequency-domain by a Fourier transform of
$S_{\mathbf{I}}(t_{3},t_{2},t_{1})$\ with respect to $t_{3}$ and $t_{1}$,
holding $t_{2}$\ fixed%

\begin{equation}
S_{\mathbf{I}}^{(3)}(\Omega_{3},t_{2},\Omega_{1})\equiv\int\int dt_{3}%
dt_{1}S_{\mathbf{I}}(t_{3},t_{2},t_{1})e^{i\Omega_{3}t_{3}}e^{i\Omega_{1}%
t_{1}}.\label{SI2D0}%
\end{equation}
This gives%

\begin{align}
S_{\mathbf{I}}^{(3)}(\Omega_{3},t_{2},\Omega_{1})  &  =\sum_{e,e^{\prime}%
}\frac{e^{-\Gamma_{gg}t_{2}}(\mathbf{e}_{1}\cdot\boldsymbol{\mu}_{ge^{\prime}%
})(\mathbf{e}_{2}\cdot\boldsymbol{\mu}_{e^{\prime}g})(\mathbf{e}_{4}%
\cdot\boldsymbol{\mu}_{ge})(\mathbf{e}_{3}\cdot\boldsymbol{\mu}_{eg}%
)\mathcal{E}_{3}^{+}\mathcal{E}_{2}^{+}\mathcal{E}_{1}^{-}}{[\Omega_{3}%
-\omega_{eg}-\omega_{1}+\omega_{2}+\omega_{3}+i\Gamma_{eg}][\Omega_{1}%
+\omega_{e^{\prime}g}-\omega_{1}+i\Gamma_{e^{\prime}g}]}\nonumber\\
&  +\sum_{e,e^{\prime}}\frac{e^{-(i\omega_{ee^{\prime}}+\Gamma_{ee^{\prime}%
})t_{2}}(\mathbf{e}_{1}\cdot\boldsymbol{\mu}_{ge^{\prime}})(\mathbf{e}%
_{3}\cdot\boldsymbol{\mu}_{e^{\prime}g})(\mathbf{e}_{4}\cdot\boldsymbol{\mu
}_{ge})(\mathbf{e}_{2}\cdot\boldsymbol{\mu}_{eg})\mathcal{E}_{3}%
^{+}\mathcal{E}_{2}^{+}\mathcal{E}_{1}^{-}}{[\Omega_{3}-\omega_{eg}-\omega
_{1}+\omega_{2}+\omega_{3}+i\Gamma_{eg}][\Omega_{1}+\omega_{e^{\prime}%
g}-\omega_{1}+i\Gamma_{e^{\prime}g}]}\nonumber\\
&  -\sum_{e,e^{\prime}}\sum_{f}\frac{e^{-(i\omega_{ee^{\prime}}+\Gamma
_{ee^{\prime}})t_{2}}(\mathbf{e}_{1}\cdot\boldsymbol{\mu}_{ge^{\prime}%
})(\mathbf{e}_{4}\cdot\boldsymbol{\mu}_{e^{\prime}f})(\mathbf{e}_{3}%
\cdot\boldsymbol{\mu}_{fe})(\mathbf{e}_{2}\cdot\boldsymbol{\mu}_{eg}%
)\mathcal{E}_{3}^{+}\mathcal{E}_{2}^{+}\mathcal{E}_{1}^{-}}{[\Omega_{3}%
-\omega_{fe^{\prime}}-\omega_{1}+\omega_{2}+\omega_{3}+i\Gamma_{fe^{\prime}%
}][\Omega_{1}+\omega_{e^{\prime}g}-\omega_{1}+i\Gamma_{e^{\prime}g}%
]}.\label{SI2D}%
\end{align}

Eq. (\ref{SI2D}) has various diagonal peaks ($\Omega_{1}=\Omega_{3}$) and
cross-peaks ($\Omega_{1}\neq\Omega_{3}$). \ The relative contributions of
different terms may be controlled by the carrier frequencies, $\omega_{1}$,
$\omega_{2}$, and $\omega_{3}$. \ Spreading the signal in an extra frequency
dimension enhances the resolving power of the 2DCS, compared to 1D
techniques.\cite{paper386} \ We can further improve the resolution\ by
controlling other parameters such as the pulse polarization directions,
carrier frequencies and envelopes. \ Other 2D techniques generated in
different phase-matching directions and using different pairs of time
variables (e.g. $t_{2}$\ and $t_{3}$\ )\ provide complementary
information\cite{paper386,paper487,Demirdoven} through different projections
of the response, as will be discussed in Section III. \ Closed expressions for
the other 2D signals $\mathbf{S}_{\mathbf{II}}$ and $\mathbf{S}_{\mathbf{III}%
}$ are given in Appendix A. \ 

\section{Application to III-V Semiconductor Quantum Wells}

\subsection{Single and Two-exciton Resonances}

The dipole selection rules for HH and LH\ excitons in III-V semiconductor
quantum wells are shown in Fig. 2. \ 

For GaAs, the electron effective mass $m_{e}=0.065m_{0}$ ($m_{0}$ is free
electron mass), while the in-plane LH and HH\ masses are $m_{LH}%
^{\shortparallel}=0.206m_{0}$ and $m_{HH}^{\shortparallel}=0.115m_{0}$
respectively, and perpendicular LH and HH\ masses are $m_{LH}^{\perp
}=0.094m_{0}$ and $m_{HH}^{\perp}=0.34m_{0}$ respectively. \ The allowed
transitions near the bandedge are denoted by right ($\mathbf{\sigma}^{+}$) and
left $\left(  \mathbf{\sigma}^{-}\right)  $ arrows, representing right and
left circularly polarized photons, respectively. \ The corresponding
transition dipoles $\mathbf{\mu}^{vc}$ are%

\begin{align}
\mathbf{\mu}^{v_{1}c_{1}}  &  =\frac{1}{\sqrt{2}}\mu_{0}\mathbf{\sigma}%
^{+}=\frac{1}{\sqrt{2}}\mu_{0}\left(  \mathbf{\hat{x}+}i\mathbf{\hat{y}}%
\right)  ,\label{dipole1}\\
\mathbf{\mu}^{v_{2}c_{2}}  &  =\frac{1}{\sqrt{2}}\mu_{0}\mathbf{\sigma}%
^{-}=\frac{1}{\sqrt{2}}\mu_{0}\left(  \mathbf{\hat{x}-}i\mathbf{\hat{y}}%
\right)  ,\\
\mathbf{\mu}^{v_{3}c_{2}}  &  =\frac{1}{\sqrt{6}}\mu_{0}\mathbf{\sigma}%
^{+}=\frac{1}{\sqrt{6}}\mu_{0}\left(  \mathbf{\hat{x}+}i\mathbf{\hat{y}}%
\right)  ,\\
\mathbf{\mu}^{v_{4}c_{1}}  &  =\frac{1}{\sqrt{6}}\mu_{0}\mathbf{\sigma}%
^{-}=\frac{1}{\sqrt{6}}\mu_{0}\left(  \mathbf{\hat{x}-}i\mathbf{\hat{y}}%
\right)  ,\label{dipole4}%
\end{align}
where $\mathbf{\hat{x}}$ and $\mathbf{\hat{y}}$ are unit vectors and $\mu_{0}
$ is the modulus of the transition dipole. \ 

We denote each type of exciton by the (hole, electron) spin values involved in
the valence to conduction band transition. \ According to Fig. 2, there are
two types of HH excitons, $\left(  -\frac{3}{2},-\frac{1}{2}\right)  $ and
$\left(  \frac{3}{2},\frac{1}{2}\right)  $, and two types of LH excitons,
$\left(  -\frac{1}{2},\text{ }\frac{1}{2}\right)  $, $\left(  \frac{1}%
{2},-\frac{1}{2}\right)  $. \ The system has 10 types of two-excitons: three
made of two single HH\ excitons are denoted $f_{H}$, three made of two single
LH excitons ( $f_{L}$), and four made of one LH exciton and one HH\ exciton
($f_{M}$, mixed). \ Two electrons (two holes ) belonging to different bands
generate a bound two-exciton

[1] $\left(  -\frac{3}{2},-\frac{1}{2}\right)  +\left(  \frac{3}{2},\frac
{1}{2}\right)  :$ bound $f_{H}$,

[2] $\left(  -\frac{1}{2},\frac{1}{2}\right)  +\left(  \frac{1}{2},-\frac
{1}{2}\right)  :$ bound $f_{L}$,

[3] \ $\left(  -\frac{3}{2},-\frac{1}{2}\right)  +\left(  -\frac{1}{2}%
,\frac{1}{2}\right)  :$ bound $f_{M}$,

[4] \ $\left(  \frac{3}{2},\frac{1}{2}\right)  +\left(  \frac{1}{2},-\frac
{1}{2}\right)  :$ bound $f_{M}$.

Other combinations produce unbound two-excitons

[5] $\left(  -\frac{3}{2},-\frac{1}{2}\right)  +\left(  -\frac{3}{2},-\frac
{1}{2}\right)  :$ unbound $f_{H}$,

[6] $\left(  \frac{3}{2},\frac{1}{2}\right)  +\left(  \frac{3}{2},\frac{1}%
{2}\right)  :$ unbound $f_{H}$,

[7] $\left(  -\frac{1}{2},\frac{1}{2}\right)  +\left(  -\frac{1}{2},\frac
{1}{2}\right)  :$ unbound $f_{L}$,

[8] $\left(  \frac{1}{2},-\frac{1}{2}\right)  +\left(  \frac{1}{2},-\frac
{1}{2}\right)  :$ unbound $f_{L}$. \ 

The other possible two-excitons $\left(  -\frac{3}{2},-\frac{1}{2}\right)  $
$+ $ $\left(  \frac{1}{2},-\frac{1}{2}\right)  $ and $\left(  \frac{3}%
{2},\frac{1}{2}\right)  $ $+$ $\left(  -\frac{1}{2},\frac{1}{2}\right)  $, are
expected to form unbound mixed two-exciton $f_{M}$ because the electron motion
dominates the internal motion of a two-exciton when the holes are much heavier
than the electrons. \ 

In Fig. 3, we show the level-scheme that includes the single-exciton and the
four bound two-exciton transitions.\cite{Wagner,Bott} \ The unbound
transitions, neglected in Fig. 3 for clarity, will be included in the
numerical calculations reported in Sec. V. \ 

The exciton to two-exciton transition dipoles are collectively denoted by
$\mathbf{\mu}^{ef}$, where $e$ can be either $e_{H}\left(  e_{H}^{\prime
}\right)  $ or $e_{L}(e_{L}^{\prime})$ and $f$ can be either $f_{H}$, $f_{L}$
or $f_{M}$. \ When all pulses are either $\sigma^{+}$or $\sigma^{-}%
$\ polarized, the model of Fig. 3 reduces to two coupled two-level systems
(dimer) whose 2DCS was studied in detail.\cite{Piryatinski} \ 

\subsection{Feynman diagrams for 2DCS}

We first examine the Liouville space pathways\cite{paper386,298} for the
$\mathbf{S}_{\mathbf{I}}$ technique. \ Starting with diagram (i) in Fig. 1 we
let the states $e$ and $e^{\prime}$ assume all possible single-exciton states,
as shown in Fig. 3. \ This yields the four Feynman diagrams (ia), (ib), (ic)
and (id) depicted in Fig. 4. \ 

The corresponding two-dimensional correlation spectrum is shown at the bottom
of Fig. 4. \ The diagonal peak (ia) comes from HH\ excitons. (ib) is similarly
a diagonal peak for LH\ excitons. \ The cross peak (ic) describes a pathway
with LH excitons during $t_{1}$\ and HH\ excitons during $t_{3}$. \ (id) is a
second cross peak representing HH excitons during $t_{1}$\ and LH\ excitons
during $t_{3}$. \ Diagram (ii) of Fig. 1 differs from (i) only during the
$t_{2}$\ period where the system is in the excitonic rather than the ground
state. \ Consequently the corresponding peaks (iia), (iib), (iic) and (iid) in
the ($\Omega_{3}$, $\Omega_{1}$) correlation plot which show the evolution
during $t_{1}$\ and $t_{3}$\ are the same as (ia), (ib), (ic) and (id), as
shown in Fig. 4. \ 

We next turn to diagram (iii)\ which involves two-exciton states. \ Solid
(open) symbols represent bound (unbound) two-excitons which are red (blue)
shifted along $\Omega_{3}$ with respect to the single-exciton peaks. \ Running
the states $e$, $e^{\prime}$ and $f$ over all possible values results in the
six diagrams shown Fig. 5. \ The corresponding 2D spectrum is shown at the
bottom of Fig. 5. \ In the total 2D spectrum (Fig. 6(a)) which combines all
spectra of Fig. 4 and Fig. 5, the two-exciton contributions (iiia) to (iiid)
appear near the four single-exciton peaks of Fig. 4 (ovals and circles).
\ There are four pathways leading to $f_{M}$, one to $f_{H}$\ and one to
$f_{L}$ respectively. \ The two-exciton peaks are either red or blue shifted
along $\Omega_{3}$\ relative to the major diagonal peaks [$(e_{H},-e_{H})$,
$(e_{L},-e_{L})$], or cross peaks [$(e_{H},-e_{L})$, $(e_{L},-e_{H})$]. \ 

To illustrate the merits of 2DCS, we compare the 2D spectra of Fig. 6(a) with
a traditional spectrally resolved FWM (1DFWM) shown in Fig. 6(b). \ Using the
current notation, the 1DFWM is given by $\mathbf{S}_{\mathbf{I}}\left(
\Omega_{3},t_{2},t_{1}\right)  $, where the two pulse-delays $t_{1}$ and
$t_{2}$ are held fixed and $\Omega_{3}$\ is varied. \ The peak denoted HH
comes from pathways (ia), (ic) and (iia), (iic). \ The LH peak originates from
the other four pathways (ib), (id) and (iib), (iid). \ In the 2D spectra (Fig.
6(a)), (iia) is separated from (iic) and (ia) is separated from (ic) along the
$\Omega_{1}$ axis. \ We find three two-exciton contributions to the 1DFWM in
Fig. 6(b) in each side of the single-exciton peak HH,\ and three in each side
of the LH peak. \ These contributions are very close and poorly resolved in
the 1D plot because the typical two-exciton binding energies in the III-V
quantum wells are 1 to 2 meV. \ Between $e_{H}$ and $e_{L}$ (3.8 meV for the
quantum well considered), there are six types of two-excitons congested in the
1D plot. \ In contrast, the bound mixed two-excitons (e.g. solid trapezoid and
square) and bound HH two-exciton (e.g. solid triangle) are well separated
along $\Omega_{1}$ in the 2D plot in Fig. 6(a). \ Moreover, additional
separation can be obtained by different projections of the $\mathbf{S}%
_{\mathbf{I}}$ signal. \ For example, the overlapping mixed two-exciton
contributions such as square and trapezoid in Fig. 6(a) (displaced for
clarity) can be separated if we plot the 2D spectra $\mathbf{S}_{\mathbf{I}%
}\left(  \Omega_{3},\Omega_{2},t_{1}\right)  $, rather than the $\mathbf{S}%
_{\mathbf{I}}\left(  \Omega_{3},t_{2},-\Omega_{1}\right)  $ in Fig. 6(a) (see
details in Appendix B).

To summarize this section, we have demonstrated how 2DCS can separate
different pathways. \ In Fig. 6(a), coherences among LH and HH\ excitons
correspond to the two major cross peaks [($e_{H},-e_{L}$), ($e_{L},-e_{H}$)]
while the many-body two-exciton correlations correspond to the weak peaks
around the four major peaks. \ 2DCS with different choice of variables or
along different phase-matching directions provide complementary information. \ 

\section{Hamiltonian and the Nonlinear Exciton Equations}

The sum-over-states expressions for 2D signals given in the previous section
provide an intuitive tool for analyzing 2D techniques. \ Calculating the
signals using these expressions requires the single-exciton and two-exciton
states and their transition dipoles and relies on the impulsive-pulse
assumption. \ This is easily done for Frenkel excitons. \ An alternative way
to proceed which is more practical for Wannier excitons is to employ the
Nonlinear Excitonic Equations (NEE)
\cite{paper487,spano1,spano2,Leegwater,paper211} or the equivalent Dynamics
Controlled Truncation (DCT) formalism
\cite{AxtRMP,chernyak,Axt94,Axt96,Lindberg1} to account for the many-body
interactions beyond the Hartree-Fock level.
\cite{Mayer2,Feuerbacher,Mayer,Mayer1,Bar-ad,lovering,Bott1,Bigot,Bigot2,Wang,Rappen,Albrecht,Schafer,Kner,Jahnke}%
\ \ To close the infinite hierarchy of dynamic variables, the equations of
motion are truncated according to the desired order of the laser
field.\cite{AxtRMP,spano1,spano2,Axt94} \ 2D signals may be calculated by
solving the NEE using nonequilibrium Green's functions and the two-exciton
scattering matrix\cite{chernyak,paper487,Mukamel1}. \ However, in this paper
we use direct integration of the equations of motion.\ \ 

Calculating the 2DCS of semiconductor quantum wells where excitons are
spatially confined in two-dimensions requires an intensive numerical effort.
\ Most computational work on the nonlinear optical response of two-exciton
correlations focused on HH\ excitons. \ Our computation time is around $300$
times higher than the corresponding 1D FWM calculation. \ To make these
calculations more tractable, we shall use a multi-band one-dimensional
tight-binding model
\cite{Banyai1,Sieh,Meier2,Euteneuer,Finger,Bott1,Brinkamann,Weiser} to
describe the excitons and two-excitons of a single quantum well. \ This model
can reproduce many spectroscopic observables in quantum-wells such as the
signs of energy shifts, bleaching and induced absorption and often even their
relative strengths, and the dependence on the polarization directions of the
incident pulses.\cite{Meierrecentbook} \ Comparison of the one-dimensional
tight-binding with two-dimensional models can be found in Ref.\cite{Sieh}.
\ Other models have been successfully used as well.\cite{Osterich I,Maialle}

We start with the multi-band tight-binding Hamiltonian is,\cite{Banyai1,Sieh,Meier2,Finger,Bott1,Brinkamann,Weiser}%

\begin{equation}
H=H_{K}+H_{C}+H_{I},\label{-1}%
\end{equation}
where%
\begin{equation}
H_{K}=\sum_{ijc}T_{ij}^{c}a_{i}^{c\dagger}a_{j}^{c}+\sum_{ijv}T_{ij}^{v}%
a_{i}^{v\dagger}a_{j}^{v}\label{0}%
\end{equation}
describes free band motion. $\ a_{i}^{c\dagger}(a_{i}^{c})$ are creation
(annihilation) Fermi operators of electrons in site $i$ from the conduction
band $c$ and $a_{i}^{v\dagger}(a_{i}^{v})$ are creation (annihilation) Fermi
operators of holes in site $i$ from the valence band $v$. \ The diagonal
elements $T_{ii}^{c,v}$ describe the site energies for the electrons (holes)
in the conduction (valence) band while the off-diagonal elements $T_{i\neq
j}^{c,v}$ represent the couplings between different sites. \ We adopt the
nearest-neighbor tight-binding approximation for the electronic coupling,
\textit{i.e.} $T_{ij}^{c,v}=0$ for $\left|  i-j\right|  >1$. \ The site
energies of the electrons and holes $T_{ii}^{c,v}$ are taken to be the mid
bandgap energy, $E_{g}$. \ 

The Coulomb term $H_{C}$ in Eq. (\ref{-1}) has the monopole-monopole form
\[
H_{C}=\sum_{ijcvc^{\prime}v^{\prime}}\left(  a_{i}^{c^{\prime}\dagger}%
a_{i}^{c^{\prime}}-a_{i}^{v^{\prime}\dagger}a_{i}^{v^{\prime}}\right)
V_{ij}\left(  a_{j}^{c\dagger}a_{j}^{c}-a_{j}^{v\dagger}a_{j}^{v}\right)  ,
\]
where the Coulomb interaction is given by%

\begin{equation}
V_{ij}=U_{0}\frac{d}{\left|  i-j\right|  d+a_{0}}.\label{mono}%
\end{equation}
This is similar to the Ohno coupling used the Pariser-Parr-Pople Hamiltonian
of conjugated molecules.\cite{MeierPRL,MeierPRB,ppp} \ $U_{0}$ characterizes
the interaction strength, $a_{0}$ is the spatial cutoff and $d$ is the lattice
constant. \ 

Finally, the dipole interaction with the radiation field has the form%

\[
H_{I}=-\mathbf{E}\left(  \mathbf{r,}t\right)  \mathbf{\cdot\hat{P},}%
\]
where $\mathbf{E}\left(  t,\mathbf{r}\right)  $ is given by Eq. (\ref{efield})
and $\mathbf{\hat{P}}$ is the interband polarization operator%
\begin{equation}
\mathbf{\hat{P}\equiv}\sum_{ijvc}\left[  \mathbf{\mu}_{ij}^{vc}p_{ij}%
^{vc}+c.c.\right]  .\label{interband}%
\end{equation}
$\mathbf{\mu}_{ij}^{vc}$ are interband dipoles and $p_{ij}^{vc}\equiv
a_{i}^{v}a_{j}^{c}\ $are interband coherences.\ \ The transition dipole matrix
elements $\mathbf{\mu}_{ij}^{vc}$ are defined in Eqs. (\ref{dipole1}) to
(\ref{dipole4}). \ All optical transitions are diagonal in the site indices,
$i$\ and $j$. \ For example, Eq. (\ref{dipole1}) reads
\[
\mathbf{\mu}_{ij}^{v_{1}c_{1}}=\frac{1}{\sqrt{2}}\mu_{0}\mathbf{\sigma}%
^{+}\delta_{ij}=\frac{1}{\sqrt{2}}\mu_{0}\delta_{ij}\left(  \mathbf{\hat{x}%
+}i\mathbf{\hat{y}}\right)  .
\]

To make the numerical calculation tractable, we neglect the exciton population
dynamics and work in the coherent limit where we consider only two types of
density matrices involving single-exciton and two-exciton
respectively.\cite{AxtRMP} To third order in the radiation field, the
equations of motion for the first type of density matrix $p_{ij}^{vc}$ are \cite{Banyai1,Sieh,Meier2,Finger,Bott1,Brinkamann,Weiser}%

\begin{align}
&  -i\frac{\partial}{\partial t}p_{ij}^{vc}-\frac{i}{t_{ex}}p_{ij}%
^{vc}\nonumber\\
&  =-\sum_{n}T_{jn}^{c}p_{in}^{vc}-\sum_{m}T_{mi}^{v}p_{mj}^{vc}+V_{ij}%
p_{ij}^{vc}\nonumber\\
&  +\sum_{klv^{\prime}c^{\prime}}\left(  V_{kj}-V_{ki}-V_{lj}+V_{li}\right)
\left[  \left(  p_{lk}^{v^{\prime}c^{\prime}}\right)  ^{\ast}p_{lj}%
^{v^{\prime}c}p_{ik}^{vc^{\prime}}\right. \nonumber\\
&  \left.  -\left(  p_{lk}^{v^{\prime}c^{\prime}}\right)  ^{\ast}%
p_{lk}^{v^{\prime}c^{\prime}}p_{ij}^{vc}-\left(  p_{lk}^{v^{\prime}c^{\prime}%
}\right)  ^{\ast}B_{lkij}^{v^{\prime}c^{\prime}vc} \right] \nonumber\\
&  +\mathbf{E}\left(  t\right)  \cdot\left[  \left(  \mathbf{\mu}_{ij}%
^{vc}\right)  ^{\ast}-\sum_{klv^{\prime}c^{\prime}}\left(  \mathbf{\mu}%
_{il}^{vc^{\prime}}\right)  ^{\ast}\left(  p_{kl}^{v^{\prime}c^{\prime}%
}\right)  ^{\ast}p_{kj}^{v^{\prime}c}\right. \nonumber\\
&  \left.  +\sum_{klv^{\prime}c^{\prime}}\left(  \mathbf{\mu}_{lj}^{v^{\prime
}c}\right)  ^{\ast}\left(  p_{lk}^{v^{\prime}c^{\prime}}\right)  ^{\ast}%
p_{ik}^{vc^{\prime}}\right]  ,\label{single}%
\end{align}
where $t_{ex}$ describes exciton dephasing time. $\ B_{lkij}^{v^{\prime
}c^{\prime}vc}\equiv a_{l}^{v^{\prime}}a_{k}^{c^{\prime}}a_{i}^{v}a_{j}^{c}$
is a two-exciton operator whose equation of motion is%

\begin{align}
&  -i\frac{\partial}{\partial t}B_{lkij}^{v^{\prime}c^{\prime}vc}-\frac
{i}{t_{bi}}B_{lkij}^{v^{\prime}c^{\prime}vc}\nonumber\\
&  =-\sum_{m}\left(  T_{jm}^{c}B_{lkim}^{v^{\prime}c^{\prime}vc}+T_{mi}%
^{v}B_{lkmj}^{v^{\prime}c^{\prime}vc}\right. \nonumber\\
&  +\left.  T_{km}^{c}B_{lmij}^{v^{\prime}c^{\prime}vc}+T_{ml}^{v}%
B_{mkij}^{v^{\prime}c^{\prime}vc}\right) \nonumber\\
&  +\left(  V_{lk}+V_{lj}+V_{ik}+V_{ij}-V_{li}-V_{kj}\right)  B_{lkij}%
^{v^{\prime}c^{\prime}vc}\nonumber\\
&  -\left(  V_{lk}+V_{ij}-V_{li}-V_{kj}\right)  p_{ik}^{vc^{\prime}}%
p_{lj}^{v^{\prime}c}\nonumber\\
&  +\left(  V_{ik}+V_{lj}-V_{li}-V_{kj}\right)  p_{lk}^{v^{\prime}c^{\prime}%
}p_{ij}^{vc},\label{double}%
\end{align}
where $t_{bi}$ is two-exciton dephasing time. \ 

The total interband polarization is given by Eq. (\ref{interband}) where
$p_{ij}^{vc}$\ is obtained by solving Eqs. (\ref{single}) and (\ref{double}).
\ To single-out a given FWM signal, we must keep track of the spatial Fourier
components of the interband polarization contributing to different 2D signals
($\mathbf{S}_{\mathbf{I}}$, $\mathbf{S}_{\mathbf{II}}$ and $\mathbf{S}%
_{\mathbf{III}}$). \ For example, to calculate $\mathbf{S}_{\mathbf{I}}$ we
need the $\mathbf{k}_{\mathbf{I}}$ component of the interband coherences,
$p_{ij}^{vc:\left[  \mathbf{k}_{\mathbf{I}}\right]  }$. \ The
positive-frequency component (first term of Eq. (\ref{interband})) of the
total interband polarization field in this direction is thus given by
\begin{equation}
\mathbf{P}^{\left[  \mathbf{k}_{\mathbf{I}}\right]  }(t_{3},t_{2}%
,t_{1},t)\equiv\sum_{ijvc}\mathbf{\mu}_{ij}^{vc}p_{ij}^{vc:\left[
\mathbf{k}_{\mathbf{I}}\right]  }(t_{3},t_{2},t_{1},t).\label{interpol}%
\end{equation}
The $\mathbf{S}_{\mathbf{I}}$ signal (Eq. (\ref{SI2D0})) is finally given by
\begin{align}
\mathbf{S}_{\mathbf{I}}(\Omega_{3},t_{2},\Omega_{1})  &  \equiv\int
\int\mathbf{P}^{\left[  \mathbf{k}_{\mathbf{I}}\right]  }(t_{3},t_{2}%
,t_{1},t)e^{i\Omega_{1}t_{1}}e^{i\Omega_{3}t_{3}}dt_{1}dt_{3}\nonumber\\
&  =\int\int\sum_{ijvc}\mathbf{\mu}_{ij}^{vc}p_{ij}^{vc:\left[  \mathbf{k}%
_{\mathbf{I}}\right]  }(t_{3},t_{2},t_{1},t)e^{i\Omega_{1}t_{1}}e^{i\Omega
_{3}t_{3}}dt_{1}dt_{3},\label{2DD}%
\end{align}
where $p_{ij}^{vc:\left[  \mathbf{k}_{\mathbf{I}}\right]  }$ depend on
$t_{3},t_{2},t_{1}$ and $t$ through Eq. (\ref{single}), as well as pulse
envelopes and carrier frequencies (Eq. (\ref{efield})). \ $p_{ij}^{vc:\left[
\mathbf{k}_{\mathbf{I}}\right]  }$ are calculated by expanding Eqs.
(\ref{single}) and (\ref{double}) in the various wavevector Fourier
components.\cite{Lindberg92} \ We have only calculated the signals for
$t_{2}=0$, where the third and second pulses coincide. \ The order by order
expansion of Eqs. (\ref{single}) and (\ref{double}) to yield the signal is
given in Appendix C. \ 

To obtain the $\mathbf{S}_{\mathbf{I}}$ signal, we have solved Eqs. (\ref{5}),
(\ref{first})\ to (\ref{fourth}) for different combinations of $(t_{3},t_{1}%
)$. \ The other signals ($\mathbf{S}_{\mathbf{II}}$ and $\mathbf{S}%
_{\mathbf{III}}$) can be calculated by deriving different sets of coupled
nonlinear equations from Eqs. (\ref{single}) and (\ref{double}) for the
relevant spatial Fourier components. \ 

\section{Numerical Results}

We have employed the 1D tight-binding model to calculate the 2DCS signal from
a 20nm GaAs/Al$_{0.3}$Ga$_{0.7}$As single quantum well (SQW).\cite{Finger}
\ For the site energies and carrier coupling energies defined in (\ref{0}), we
used $T_{i\neq j}^{c}=8$ meV, $T_{i\neq j}^{v=1,2}=4.75$ meV (HH band) and
$T_{i\neq j}^{v=3,4}=2.52$ meV (LH band) to account for the in-plane
dispersion of the valence-band structure in the quantum
well.\cite{Meier2,Koch} \ The site energies $T_{i=j}^{c}$ and $T_{i=j}%
^{v=1,2}$ are taken to be half of the bandgap, $E_{g}$. \ We used Gaussian
pulse envelopes $\mathcal{E}_{j}^{\pm}\left(  t-t_{j}\right)  =\exp\left[
-\left(  t-t_{j}\right)  ^{2}/\delta_{j}^{2}\right]  $, where $\delta_{j}=0.6$
ps, corresponding to a spectral bandwidth of around 5 meV (FWHM). \ This
narrow bandwidth excludes the continuum states and allows to control the
relative strength of the LH and HH\ exciton transitions by tuning the carrier
frequency, since it is comparable to the LH\ and HH\ splitting (3.8 meV) of
the SQW sample. \ The pulse power spectra and the linear absorption are shown
in Fig. 7. Calculational details are presented in Appendix D. \ 

Because the oscillator strengths of LH excitons are approximately one third of
those of HH\ excitons, we have tuned the carrier frequencies of the three
optical pulses $\omega_{1}=\omega_{2}=\omega_{3}\equiv\omega_{c}$ to
$\omega_{c}=3$ meV (Fig. 7, dotted), which enhances the LH exciton
transitions. \ The frequency scale in Fig. 7 and all remaining figures are
relative to the HH exciton resonance which is 35 meV below the bandgap. \ 

To calculate the 2D spectra, we first computed $\mathbf{P}^{\left[
\mathbf{k}_{\mathbf{I}}\right]  }$ for fixed $t_{1}$ from $t_{3}=0$ to 32 ps
with 660 time grid points by solving Eqs. (\ref{5}) to (\ref{fourth}). \ These
calculations were then repeated by varying $t_{1}$ from 2 ps to 18 ps on a
330-point grid. \ The computation time scales as the number of equations, i.e.
64$\cdot N^{4}$, where $N$\ is the number of sites. \ We chose $t_{ex}=2 $ ps
and $t_{bi}=1$ ps, rather than $t_{ex}=4$ ps and $t_{bi}=2$ ps for the
dephasing times of excitons and two-excitons as in Ref. \cite{Finger}. The
numerical simulations converge better for using faster dephasing rates but at
the expense of the spectral resolution. \ The current values of dephasing
times (2ps/1ps for excitons and two-excitons) do not show two-exciton peaks
but merely show shoulders. \ In all calculations, we employed periodic
boundary conditions for $N=10$. \ Adding more sites did not have significant
effect on our calculations. \ In Fig. 8, we show $\left|  \mathbf{S}%
_{\mathbf{I}}(\Omega_{3},t_{2},t_{1},)\right|  $ for $t_{1}=2$ ps\ for
different basis size $N=$10, 12, 14 and 16. \ The main features of the spectra
have converged at $N=10$. \ Higher energy (%
$>$
8 meV) peaks represent continuum states. \ These are not converged at $N=10$.
\ However, for the narrow pulse bandwidth used (Fig. 7), the contribution of
these continuum states is very small. \ Testing the convergence is expensive.
\ A single 1DFWM calculation with 16(18) sites for current parameters takes
more than 20(32) hours on a single processor (64-bit opteron). \ We thus
tested the convergence of 1DFWM signal before launching the 2DCS calculations.
\ Fast Fourier Transform of the $660\times330$ grid of $\mathbf{P}^{\left[
\mathbf{k}_{\mathbf{I}}\right]  }(t_{3},t_{2},t_{1},t)$ with respect to
$t_{1}$ and $t_{3}$ gave $\mathbf{S}_{\mathbf{I}}(\Omega_{3},t_{2},-\Omega
_{1})$. \ The typical computational time for one 2D plot for $N=10$ is around
1000 hours on a single processor (64-bit opteron). \ 

The calculated $\left|  \mathbf{S}_{\mathbf{I}}(\Omega_{3},t_{2},-\Omega
_{1})\right|  $ signal with two co-linearly polarized pulses (hereafter
referred to as XX excitation) is shown in Fig. 9-A (log scale). \ 

The two diagonal peaks ($e_{H},-e_{H}$) and ($e_{L},-e_{L}$)\ describe
respectively the HH and LH\ excitons. \ Since the pulse is tuned closer to the
LH\ exciton resonance, the peak ($e_{L},-e_{L}$) is stronger than
($e_{H},-e_{H}$). \ The two major cross peaks ($e_{H},-e_{L}$) [pathways (ic)
and (iic)]$\ $and $(e_{L},-e_{H})\ $[pathways (id) and (iid)]$\ $describe the
coherences which lead to quantum beating among LH and HH excitons. \ These
pathways may be separated using $\mathbf{S}_{\mathbf{I}}\left(  \Omega
_{3},\Omega_{2},t_{1}\right)  $ as shown in Fig. 13, or the $\mathbf{S}%
_{\mathbf{II}}$ technique (Appendix A). \ 

To show the two-exciton signatures in Fig. 9-A, we have calculated the 2D
spectra without the two-exciton contributions (Fig. 9-B) obtained by setting
the term $B_{lkij}^{v^{\prime}c^{\prime}vc}$ in Eq. (\ref{5}) to zero.
\ Comparing Fig. 9-A and 9-B,\ we can clearly see a feature (e) in Fig. 9-A
for mixed unbound two-excitons. \ This corresponds to pathways, (iiia') and
(iiid). \ Further separation of these two pathways can be accomplished using
other techniques as described in Fig. 13 or in Appendix A. \ All the
two-exciton peaks appear as shoulders adjacent to major single-exciton peaks,
or as a broadening of the single-exciton peaks. \ 

The two-exciton feature (b) in Fig. 9-A can be attributed to the bound
LH\ two-excitons discussed in Fig. 6(a) and is visible since the optical pulse
enhances the LH excitons as compared to HH\ excitons. \ Some other two-exciton
contributions are also seen in Fig. 9-A. \ For example, a weak shoulder (c),
two shoulders (a) and (d) are seen in Fig. 9-A, but not Fig. 9-B. \ As
expected, the two contributions (a)\ and (d) become stronger as the pulses are
further tuned to the red (see Fig. 10-A). \ Finally, by comparing the 2D plots
(not shown) with and without the term including the factor $\mathbf{E}\left(
t\right)  $ in Eq. (\ref{5}), we found that the contribution from
Pauli\ blocking\cite{3rdorder} is negligible. \ 

To study the dependence of $\mathbf{S}_{\mathbf{I}}$ on the pulse polarization
configuration, we have calculated the spectra with $\sigma^{+}\sigma^{+}$
excitation. \ These spectra\ shown in Fig. 9-C\ are similar to Fig. 9-A.
\ However, the signature of bound LH\ two-exciton, the shoulder (b) in Fig.
9-A, is now absent. \ This is because, according to the selection rules shown
in Fig. 3, $\sigma^{+}\sigma^{+}$ excitation cannot generate either bound
HH\ two-excitons or bound LH\ two-excitons. \ 

Fig. 10-A shows the 2D spectrum when the carrier frequency is tuned to
$\omega_{c}=2$ meV (Fig. 7, dashed dot). \ The 2D spectra calculated without
two-exciton contributions is shown in Fig. 10-B. \ 

The effect of tuning is demonstrated by comparing Fig. 10-A ($\omega_{c}=2$
meV) and Fig. 9-A ($\omega_{c}=3$ meV). \ The strongest exciton peak
$(e_{L},-e_{L})$\ in Fig. 9-A becomes the weakest in Fig. 10-A, while the
weakest exciton peak $(e_{H},-e_{H})$\ in Fig. 9-A becomes the strongest in
Fig. 10-A. \ Furthermore, because the LH\ excitons are no longer enhanced for
$\omega_{c}=2$ meV, the signature of the LH\ bound two-excitons, the shoulder
(b) in Fig. 9-A, is absent in Fig. 10-A. \ The two-exciton features (c) and
(e) in Fig. 9-A become weaker in Fig. 10-A. \ However, the two-exciton
contributions (a) and (d) in Fig. 10-A, become stronger. \ 

By further red-shifting the pulses carrier frequency to $\omega_{c}=0$ meV
(short dot in Fig. 7) and using cross-linear (XY) polarization (first pulse is
polarized in X direction and the second in Y direction), we obtain the 2D
spectra in Fig. 11-A. \ For this excitation condition, the pulses are resonant
with the HH excitons, and LH excitons are barely excited. \ Due to the narrow
spectral pulse bandwidth, the LH and HH excitons are weakly coupled and the
$(e_{L},-e_{H})\ $and ($e_{H},-e_{L}$) peaks are very weak. \ For two-exciton
contributions, we can clearly see a shoulder (a) for bound HH two-excitons and
a broadening (b) for unbound HH\ two-excitons, as compared to Fig. 11-B where
two-exciton contributions are excluded. \ The two-exciton contributions are
more evident for XY excitation due to strong cancellations of the excitonic
signal components.\cite{Weiser} \ Experimental measurements of the XY
configuration also show the elongation along the $\Omega_{3}$ axis, consistent
with the calculated spectrum. \ 

We have further calculated the phase-sensitive 2D spectra rather than their
modulus. \ Re($\mathbf{S}_{\mathbf{I}}(\Omega_{3},t_{2},-\Omega_{1})$),
calculated with $\omega_{c}=0$ meV (XX), is shown in Fig. 12-A. \ The peak
$(e_{H},-e_{H})$ resembles that experimentally observed (peak RA of Fig. 2 in
Ref. \cite{Cundiff06PRL}). \ The 2D spectra in Fig. 12 is plotted in the same
way as in Ref.\cite{Cundiff06PRL} \ 

The sum-over-states expression for $\mathbf{S}_{\mathbf{I}}$ (Eq.
(\ref{SI2D})) may be used to interpret the phase-sensitive spectra in Fig.
12-A. \ For the two-pulse scheme ($t_{2}=0$) and colinear excitation, we
define the real parameter $C\equiv e^{-\Gamma_{gg}t_{2}}(\mathbf{e}_{1}%
\cdot\boldsymbol{\mu}_{ge^{\prime}})(\mathbf{e}_{2}\cdot\boldsymbol{\mu
}_{e^{\prime}g})(\mathbf{e}_{4}\cdot\boldsymbol{\mu}_{ge})(\mathbf{e}_{3}%
\cdot\boldsymbol{\mu}_{eg})\mathcal{E}_{3}^{+}\mathcal{E}_{2}^{+}%
\mathcal{E}_{1}^{-}$, where we assume the $\mathcal{E}_{3}^{+}\mathcal{E}%
_{2}^{+}\mathcal{E}_{1}^{-}$ is real. \ Moreover, for co-linear excitation,
the two-exciton contributions from the third term in Eq. (\ref{SI2D}) are
small and will be neglected. \ We further assume the carrier frequencies
$\omega_{1}=\omega_{2}=\omega_{3}\equiv\omega_{c}$. \ Eq. (\ref{SI2D}) then gives%

\begin{align}
S_{I}^{(3)}(\Omega_{3},t_{2},-\Omega_{1})  &  =\sum_{e,e^{\prime}}\frac
{C}{[\Omega_{3}-\tilde{\omega}_{eg}+i\Gamma_{eg}][\Omega_{1}+\tilde{\omega
}_{e^{\prime}g}+i\Gamma_{e^{\prime}g}]}\nonumber\\
&  +\sum_{e,e^{\prime}}\frac{C}{[\Omega_{3}-\tilde{\omega}_{eg}+i\Gamma
_{eg}][\Omega_{1}+\tilde{\omega}_{e^{\prime}g}+i\Gamma_{e^{\prime}g}%
]}\nonumber\\
&  =\sum_{e,e^{\prime}}\frac{2C}{[\Omega_{3}-\tilde{\omega}_{eg}+i\Gamma
_{eg}][\Omega_{1}+\tilde{\omega}_{e^{\prime}g}+i\Gamma_{e^{\prime}g}%
]},\label{SI2D1}%
\end{align}
where $\tilde{\omega}_{eg}\equiv\omega_{eg}-\omega_{c}$ and $\tilde{\omega
}_{e^{\prime}g}\equiv\omega_{e^{\prime}g}-\omega_{c}$. \ We rewrite Eq.
(\ref{SI2D1}) as%

\begin{equation}
S_{I}^{(3)}(\Omega_{3},t_{2},-\Omega_{1})=\sum_{e,e^{\prime}}\frac{2C\left[
\left(  \Omega_{3}-\tilde{\omega}_{eg}\right)  -i\Gamma_{eg}\right]  \left[
\left(  \Omega_{1}+\tilde{\omega}_{e^{\prime}g}\right)  -i\Gamma_{e^{\prime}%
g}\right]  }{\left[  \left(  \Omega_{3}-\tilde{\omega}_{eg}\right)
^{2}+\Gamma_{eg}^{2}\right]  \left[  \left(  \Omega_{1}+\tilde{\omega
}_{e^{\prime}g}\right)  ^{2}+\Gamma_{e^{\prime}g}^{2}\right]  }\label{SI2D2}%
\end{equation}
and thus
\begin{equation}
\operatorname{Re}\left[  S_{I}^{(3)}(\Omega_{3},t_{2},-\Omega_{1})\right]
=\sum_{e,e^{\prime}}\frac{2C\left(  \Omega_{3}-\tilde{\omega}_{eg}\right)
\left(  \Omega_{1}+\tilde{\omega}_{e^{\prime}g}\right)  -2C\Gamma_{eg}%
\Gamma_{e^{\prime}g}}{\left[  \left(  \Omega_{3}-\tilde{\omega}_{eg}\right)
^{2}+\Gamma_{eg}^{2}\right]  \left[  \left(  \Omega_{1}+\tilde{\omega
}_{e^{\prime}g}\right)  ^{2}+\Gamma_{e^{\prime}g}^{2}\right]  }.\label{SI2D3}%
\end{equation}
For the excitation of only HH exciton as in Fig. 12-A, Eq. (\ref{SI2D3}) is
simplified as
\begin{equation}
\operatorname{Re}\left[  S_{I}^{(3)}(\Omega_{3},t_{2},-\Omega_{1})\right]
=\frac{2C\left[  \left(  \Omega_{3}-\tilde{\omega}_{HH}\right)  \left(
\Omega_{1}+\tilde{\omega}_{HH}\right)  -\Gamma_{HH}^{2}\right]  }{\left[
\left(  \Omega_{3}-\tilde{\omega}_{HH}\right)  ^{2}+\Gamma_{HH}^{2}\right]
\left[  \left(  \Omega_{1}+\tilde{\omega}_{HH}\right)  ^{2}+\Gamma_{HH}%
^{2}\right]  },\label{SI2D4}%
\end{equation}
where $\tilde{\omega}_{HH}$ is the HH\ exciton transition frequency relative
to the carrier frequency and $\Gamma_{HH}$ is the exciton dephasing rate. \ 

Eq. (\ref{SI2D4}) accounts for the main features of Fig. 12-A. \ The
denominator gives the resonance peak at $\Omega_{3}=\tilde{\omega}_{HH}$ and
$\Omega_{1}=-\tilde{\omega}_{HH}$, while the numerator gives the
positive/negative feature. \ When neglecting $\Gamma_{HH}^{2}$,\ we see that
if $\Omega_{3}$ goes from below to above $\tilde{\omega}_{HH}$, the signal
will change sign for fixed $\Omega_{1}$. \ Similarly, when $\Omega_{1}$ is
scanned across -$\tilde{\omega}_{HH}$, the signal will also change sign for
fixed $\Omega_{3}$. \ In the region where the signal changes signs, steep
gradient occurs in the form of the dense contour lines in the center of the
blue/red peak shown in Fig. 12-A. \ The zero line along which the steepest
local gradient occurs must pass through the peak center $\left(  \tilde
{\omega}_{HH},-\tilde{\omega}_{HH}\right)  $ as at this point Eq.
(\ref{SI2D4}) vanishes when $\Gamma_{HH}^{2}$ is neglected. \ However, for
finite $\Gamma_{HH}^{2}$, Eq. (\ref{SI2D4}) is finite at $\left(
\tilde{\omega}_{HH},-\tilde{\omega}_{HH}\right)  $ and thus the zero contour
line with steepest local gradient, or the line passing through the center of
the negative/positive (blue/red) peak will not pass through the point $\left(
\tilde{\omega}_{HH},-\tilde{\omega}_{HH}\right)  $, which is on the diagonal
line. \ Thus, due to the finite dephasing, the peak center will be off the
diagonal line. \ Because $\Gamma_{HH}^{2}$ is dominated by Coulomb
interactions beyond first-order (carrier-carrier scattering), the asymmetry of
the peaks and their offset from the diagonal lines may provide insights on
these many-body correlations. \ The above analysis applies only to the
excitation of a single level, e.g. HH excitons. \ The picture will become more
complicated if a multi-level system is excited. \ 

The calculated 2D spectra (real part) with carrier frequency $\omega_{c}=2$
meV (XX) shown in Fig. 12-B share some similarities with recent
experiments\cite{Cundiff06PRL}, especially for the main peak $(e_{H},-e_{H})$.
\ However, it is difficult to produce 2D spectra with all four peaks observed
in the experiment. \ In our calculations, we use a different SQW as compared
to the experiment. \ A two-pulse scheme is employed in our simulations while
the experiment was done with three pulses. \ In addition, the 2D spectra are
very sensitive to the carrier frequency of the narrow-spectra optical pulses
and to the parameters such as site energies, hopping and dephasing rates for
the HH and LH excitons. \ For example, by tuning the pulse frequency to the
blue ($\omega_{c}=3$ meV (XX)), the real signal (Fig. 12-C)\ changes
considerably compared to Fig. 12-B. \ The experimental data also show strong
dependences on tuning of the excitation pulses as well as their
strength.\cite{Cundiff06PRL,CundiffCPL} \ Finally, incorporating the
population dynamics and continuum states will be required to fully reproduce
the experimental spectra. \ 

\section{Conclusions}

As in NMR, optical 2D techniques supplement traditional 1D techniques by
revealing microscopic couplings, coherences and many-body correlations.
\ Using the Liouville pathway analysis, we have predicted the exciton and
two-exciton contributions to 2DCS in a single quantum well with LH\ and HH
exciton couplings. \ Numerical simulations reproduce the major diagonal peaks
and cross-peaks in a 2D spectra. \ We also obtained several features from
two-exciton contributions. \ Different 2D techniques provide complementary
information. \ 

Some features predicted by the sum-over-states expressions, for example the
bound mixed two-excitons, are not seen in our simulations. This may be due to
the small binding energies of these bound two excitons\ and to the excitation
conditions used. \ Future extension of the current two-pulse scheme to
three-pulses\cite{Tanimura} and exploring the $t_{2}$-dependent 2D spectra
will be of interest. \ 

Acknowledgement: The support of the National Science Foundation (Grant nos.
CHE-0446555 and EEC-0406750) and the National Institutes of the Health (Grant
no. 2R01-GM59230-05) is gratefully acknowledged. \ We wish to thank Drs. T.
Meier, D. Abramavicius, M. M. Dignam for most useful discussions. \ 

\appendix

\section{The $\mathbf{S}_{\mathbf{II}}$ and $\mathbf{S}_{\mathbf{III}}$ techniques}

The response function for the $\mathbf{k}_{\mathbf{II}}=\mathbf{k}%
_{1}-\mathbf{k}_{2}+\mathbf{k}_{3}$ signal is given by%

\begin{align}
\lefteqn{R_{\mathbf{II}}^{(3)}(t_{3},t_{2},t_{1})}\label{time2}\\
&  =i^{3}\sum_{e,e^{\prime}}(\mathbf{e}_{4}\cdot\boldsymbol{\mu}_{ge^{\prime}%
})(\mathbf{e}_{3}\cdot\boldsymbol{\mu}_{e^{\prime}g})(\mathbf{e}_{2}%
\cdot\boldsymbol{\mu}_{ge})(\mathbf{e}_{1}\cdot\boldsymbol{\mu}_{eg}%
)e^{-(i\omega_{e^{\prime}g}+\Gamma_{e^{\prime}g})t_{3}-\Gamma_{gg}%
t_{2}-(i\omega_{eg}+\Gamma_{eg})t_{1}}\nonumber\\
&  +i^{3}\sum_{e,e^{\prime}}(\mathbf{e}_{2}\cdot\boldsymbol{\mu}_{ge^{\prime}%
})(\mathbf{e}_{3}\cdot\boldsymbol{\mu}_{e^{\prime}g})(\mathbf{e}_{4}%
\cdot\boldsymbol{\mu}_{ge})(\mathbf{e}_{1}\cdot\boldsymbol{\mu}_{eg}%
)e^{-(i\omega_{eg}+\Gamma_{eg})t_{3}-(i\omega_{ee^{\prime}}+\Gamma
_{ee^{\prime}})t_{2}-(i\omega_{eg}+\Gamma_{eg})t_{1}}\nonumber\\
&  -i^{3}\sum_{e,e^{\prime}}\sum_{f}(\mathbf{e}_{2}\cdot\boldsymbol{\mu
}_{ge^{\prime}})(\mathbf{e}_{4}\cdot\boldsymbol{\mu}_{e^{\prime}f}%
)(\mathbf{e}_{3}\cdot\boldsymbol{\mu}_{fe})(\mathbf{e}_{1}\cdot\boldsymbol
{\mu}_{eg})e^{-(i\omega_{fe^{\prime}}+\Gamma_{fe^{\prime}})t_{3}%
-(i\omega_{ee^{\prime}}+\Gamma_{ee^{\prime}})t_{2}-(i\omega_{eg}+\Gamma
_{eg})t_{1}}.\nonumber
\end{align}
where the three terms are represented by the Feynman diagrams (iv), (v) and
(vi) of Fig. 1 respectively. \ Substituting Eq. (\ref{time2}) into Eq.
(\ref{2}) gives time-domain signal $S_{\mathbf{II}}(t_{3},t_{2},t_{1})$. \ The
2D\ correlation plot is then obtained by Fourier transforming $S_{\mathbf{II}%
}(t_{3},t_{2},t_{1})$ with respect to $t_{1}$ and $t_{3}$, holding $t_{2}$
fixed, i.e.,
\begin{equation}
S_{\mathbf{II}}^{(3)}(\Omega_{3},t_{2},\Omega_{1})\equiv\int\int dt_{3}%
dt_{1}S_{\mathbf{II}}(t_{3},t_{2},t_{1})e^{i\Omega_{3}t_{3}}e^{i\Omega
_{1}t_{1}}.
\end{equation}
We get%

\begin{align}
&  S_{\mathbf{II}}^{(3)}(\Omega_{3},t_{2},\Omega_{1})\\
&  =\sum_{e,e^{\prime}}\frac{e^{-\Gamma_{gg}t_{2}}(\mathbf{e}_{4}%
\cdot\boldsymbol{\mu}_{ge^{\prime}})(\mathbf{e}_{3}\cdot\boldsymbol{\mu
}_{e^{\prime}g})(\mathbf{e}_{2}\cdot\boldsymbol{\mu}_{ge})(\mathbf{e}_{1}%
\cdot\boldsymbol{\mu}_{eg})\mathcal{E}_{3}^{+}\mathcal{E}_{2}^{-}%
\mathcal{E}_{1}^{+}}{[\Omega_{3}-\omega_{e^{\prime}g}+\omega_{3}-\omega
_{2}+\omega_{1}+i\Gamma_{e^{\prime}g}][\Omega_{1}-\omega_{eg}+\omega
_{1}+i\Gamma_{eg}]}\nonumber\\
&  +\sum_{e,e^{\prime}}\frac{e^{-(i\omega_{ee^{\prime}}+\Gamma_{ee^{\prime}%
})t_{2}}(\mathbf{e}_{2}\cdot\boldsymbol{\mu}_{ge^{\prime}})(\mathbf{e}%
_{3}\cdot\boldsymbol{\mu}_{e^{\prime}g})(\mathbf{e}_{4}\cdot\boldsymbol{\mu
}_{ge})(\mathbf{e}_{1}\cdot\boldsymbol{\mu}_{eg})\mathcal{E}_{3}%
^{+}\mathcal{E}_{2}^{-}\mathcal{E}_{1}^{+}}{[\Omega_{3}-\omega_{eg}+\omega
_{3}-\omega_{2}+\omega_{1}+i\Gamma_{eg}][\Omega_{1}-\omega_{eg}+\omega
_{1}+i\Gamma_{eg}]}\nonumber\\
&  -\sum_{e,e^{\prime}}\sum_{f}\frac{e^{-(i\omega_{ee^{\prime}}+\Gamma
_{ee^{\prime}})t_{2}}(\mathbf{e}_{2}\cdot\boldsymbol{\mu}_{ge^{\prime}%
})(\mathbf{e}_{4}\cdot\boldsymbol{\mu}_{e^{\prime}f})(\mathbf{e}_{3}%
\cdot\boldsymbol{\mu}_{fe})(\mathbf{e}_{1}\cdot\boldsymbol{\mu}_{eg}%
)\mathcal{E}_{3}^{+}\mathcal{E}_{2}^{-}\mathcal{E}_{1}^{+}}{[\Omega_{3}%
-\omega_{fe^{\prime}}+\omega_{3}-\omega_{2}+\omega_{1}+i\Gamma_{fe^{\prime}%
}][\Omega_{1}-\omega_{eg}+\omega_{1}+i\Gamma_{eg}]}\nonumber
\end{align}

The response function for the $\mathbf{k}_{\mathbf{III}}=\mathbf{k}%
_{1}+\mathbf{k}_{2}-\mathbf{k}_{3}$ signal is given by%

\begin{align}
&  R_{\mathbf{III}}^{(3)}(t_{3},t_{2},t_{1})\label{time3}\\
&  =i^{3}\sum_{e,e^{\prime}}\sum_{f}(\mathbf{e}_{4}\cdot\boldsymbol{\mu
}_{ge^{\prime}})(\mathbf{e}_{3}\cdot\boldsymbol{\mu}_{e^{\prime}f}%
)(\mathbf{e}_{2}\cdot\boldsymbol{\mu}_{fe})(\mathbf{e}_{1}\cdot\boldsymbol
{\mu}_{eg})e^{-(i\omega_{e^{\prime}g}+\Gamma_{e^{\prime}g})t_{3}-(i\omega
_{fg}+\Gamma_{fg})t_{2}-(i\omega_{eg}+\Gamma_{eg})t_{1}}\nonumber\\
&  -i^{3}\sum_{e,e^{\prime}}\sum_{f}(\mathbf{e}_{3}\cdot\boldsymbol{\mu
}_{ge^{\prime}})(\mathbf{e}_{4}\cdot\boldsymbol{\mu}_{e^{\prime}f}%
)(\mathbf{e}_{2}\cdot\boldsymbol{\mu}_{fe})(\mathbf{e}_{1}\cdot\boldsymbol
{\mu}_{eg})e^{-(i\omega_{fe^{\prime}}+\Gamma_{fe^{\prime}})t_{3}-(i\omega
_{fg}+\Gamma_{fg})t_{2}-(i\omega_{eg}+\Gamma_{eg})t_{1}}\nonumber
\end{align}
\ where the two terms correspond to the Feynman diagrams (vii) and (viii) in
Fig. 1. \ Substituting Eq. (\ref{time3}) into Eq. (\ref{3}) gives time-domain
signal $S_{\mathbf{III}}(t_{3},t_{2},t_{1})$. \ The 2D correlation spectra are
obtained by Fourier transform $S_{\mathbf{III}}(t_{3},t_{2},t_{1})$ with
respect to $t_{2}$ and $t_{3}$, holding $t_{1}$ fixed,%

\[
S_{\mathbf{III}}^{(3)}(\Omega_{3},\Omega_{2},t_{1})\equiv\int\int dt_{3}%
dt_{1}S_{\mathbf{III}}(t_{3},t_{2},t_{1})e^{i\Omega_{3}t_{3}}e^{i\Omega
_{2}t_{2}}.
\]
This gives
\begin{align}
&  S_{\mathbf{III}}^{(3)}(\Omega_{3},\Omega_{2},t_{1})\nonumber\\
&  =\sum_{e,e^{\prime}}\sum_{f}\frac{e^{-(i\omega_{eg}+\Gamma_{eg})t_{1}%
}(\mathbf{e}_{4}\cdot\boldsymbol{\mu}_{ge^{\prime}})(\mathbf{e}_{3}%
\cdot\boldsymbol{\mu}_{e^{\prime}f})(\mathbf{e}_{2}\cdot\boldsymbol{\mu}%
_{fe})(\mathbf{e}_{1}\cdot\boldsymbol{\mu}_{eg})\mathcal{E}_{3}^{-}%
\mathcal{E}_{2}^{+}\mathcal{E}_{1}^{+}}{[\Omega_{3}-\omega_{e^{\prime}%
g}-\omega_{3}+\omega_{2}+\omega_{1}+i\Gamma_{e^{\prime}g}][\Omega_{2}%
-\omega_{fg}+\omega_{1}+\omega_{2}+i\Gamma_{fg}]}\nonumber\\
&  -\sum_{e,e^{\prime}}\sum_{f}\frac{e^{-(i\omega_{eg}+\Gamma_{eg})t_{1}%
}(\mathbf{e}_{3}\cdot\boldsymbol{\mu}_{ge^{\prime}})(\mathbf{e}_{4}%
\cdot\boldsymbol{\mu}_{e^{\prime}f})(\mathbf{e}_{2}\cdot\boldsymbol{\mu}%
_{fe})(\mathbf{e}_{1}\cdot\boldsymbol{\mu}_{eg})\mathcal{E}_{3}^{-}%
\mathcal{E}_{2}^{+}\mathcal{E}_{1}^{+}}{[\Omega_{3}-\omega_{fe^{\prime}%
}-\omega_{3}+\omega_{2}+\omega_{1}+i\Gamma_{fe^{\prime}}][\Omega_{2}%
-\omega_{fg}+\omega_{1}+\omega_{2}+i\Gamma_{fg}]}.\label{s3}%
\end{align}

$S_{\mathbf{II}}$ and $S_{\mathbf{III}}$ techniques provide complementary
information to the $S_{\mathbf{I}}$ technique. \ $S_{\mathbf{III}}$ technique
is particular sensitive to exciton-exciton coupling. \ The 2D spectra of
techniques $S_{\mathbf{II}}$ and $S_{\mathbf{III}}$ can be similarly analyzed
as in the $S_{\mathbf{I}}$ technique. \ 

\section{Alternative 2D Projection of the $\mathbf{S}_{\mathbf{I}}$ Signal}

The bound mixed two-excitons from different pathways (e.g. solid trapezoid and
square) not resolved in Fig. 6(a) may be separated by displaying the
$\mathbf{S}_{\mathbf{I}}$ 2D spectra with different variables such as
$(\Omega_{3},\Omega_{2},t_{1})$ rather than $\left(  \Omega_{3},t_{2}%
,-\Omega_{1}\right)  $, or by employing other 2D techniques (Appendix A).
$\ $In Fig. 13 we display $\mathbf{S}_{\mathbf{I}}(\Omega_{3},\Omega_{2}%
,t_{1})$. \ Panels (a) and (b) correspond to the Feynman diagrams in Fig. 4
and Panel (c) corresponds to the Feynman diagrams in Fig. 5. \ Panel (d) shows
the total spectrum from the diagrams of Figs. 4 and 5, where the
single-exciton contributions along $\Omega_{3}$ (from Fig. 13(a) and (b)) are
represented by two open vertical ovals. \ $\mathbf{S}_{\mathbf{I}}(\Omega
_{3},\Omega_{2},t_{1})$ can resolve peaks indistinguishable in $\mathbf{S}%
_{\mathbf{I}}\left(  \Omega_{3},t_{2},-\Omega_{1}\right)  $. For example, (ic)
and (iic) (both denoted by a horizontal oval) overlap at $(e_{H},-e_{L})$\ in
Fig. 6(a). \ Similarly, (id) and (iid) (both denoted by a vertical oval)
overlap at $(e_{L},-e_{H})$. \ The two-exciton contributions denoted by a
solid trapezoid and a solid square in Fig. 6(a) also overlap. \ The same holds
for the two-exciton contributions denoted by a solid hexagon and a solid
inverted triangle in Fig. 6(a). \ However, all of these overlapping couplings
can be easily resolved by $\mathbf{S}_{\mathbf{I}}(\Omega_{3},\Omega_{2}%
,t_{1})$. \ In Fig. 13(a), pathways (ia) and (ic) both contribute to the peak
$\left(  \Omega_{3},\Omega_{2}\right)  =(e_{H},0)$ (displaced in the figure
for clarity). \ However pathway (iic) now gives a peak at $(e_{H},e_{H}%
-e_{L})$\ (Fig. 13(b)) and is thus well separated from (ic), in contrast to
$\mathbf{S}_{\mathbf{I}}\left(  \Omega_{3},t_{2},-\Omega_{1}\right)  $ (Fig.
4) where these two pathways are overlapped. \ Similarly, pathway (iid) appears
at $(e_{L},e_{L}-e_{H})$\ (Fig. 13(b)) and is thus separated from (id), which
appears at $(e_{L},0)$ in Fig. 13(a). \ Thus $\mathbf{S}_{\mathbf{I}}\left(
\Omega_{3},\Omega_{2},t_{1}\right)  $ (Fig. 13) provides a complementary
information by separating different pathways which overlap in $\mathbf{S}%
_{\mathbf{I}}\left(  \Omega_{3},t_{2},-\Omega_{1}\right)  $ (Fig. 6(a)). \ All
two-exciton contributions to Fig. 5 are plotted in Fig. 13(c). \ Fig. 13(c)
also provides complementary information to $\mathbf{S}_{\mathbf{I}}\left(
\Omega_{3},t_{2},-\Omega_{1}\right)  $ (Fig. 6(a)). \ For example, the mixed
two-excitons from two pathways (solid square and a trapezoid) overlap in Fig.
6(a) but are well separated in Fig. 13(c). \ The same holds for the solid
hexagon and solid inverted triangle. \ 

\section{Selecting Spatial Fourier Components of Interband Polarizations}

In this appendix, we present the equations of motion used for calculating FWM
signals for a two-pulse scheme ($t_{2}=0$). \ According to Eq. (\ref{single}),
the $\mathbf{k}_{\mathbf{I}}$ spatial component of interband density matrix,
$p_{ij}^{vc:\left[  \mathbf{k}_{\mathbf{I}}\right]  }$, satisfies%

\begin{align}
&  -i\frac{\partial}{\partial t}p_{ij}^{vc:\left[  -\mathbf{k}_{1}%
+\mathbf{k}_{2}+\mathbf{k}_{2}\right]  }-\frac{i}{t_{ex}}p_{ij}^{vc:\left[
-\mathbf{k}_{1}+\mathbf{k}_{2}+\mathbf{k}_{2}\right]  }\nonumber\\
&  =-\sum_{n}T_{jn}^{c}p_{in}^{vc:\left[  -\mathbf{k}_{1}+\mathbf{k}%
_{2}+\mathbf{k}_{2}\right]  }\\
&  -\sum_{m}T_{mi}^{v}p_{mj}^{vc:\left[  -\mathbf{k}_{1}+\mathbf{k}%
_{2}+\mathbf{k}_{2}\right]  }+V_{ij}p_{ij}^{vc:\left[  -\mathbf{k}%
_{1}+\mathbf{k}_{2}+\mathbf{k}_{2}\right]  }\nonumber\\
&  +\sum_{klv^{\prime}c^{\prime}}\left(  V_{kj}-V_{ki}-V_{lj}+V_{li}\right)
\nonumber\\
&  \cdot\left\{  \left[  \left(  p_{lk}^{v^{\prime}c^{\prime}}\right)  ^{\ast
}p_{lj}^{v^{\prime}c}p_{ik}^{vc^{\prime}}\right]  ^{\left[  -\mathbf{k}%
_{1}+\mathbf{k}_{2}+\mathbf{k}_{2}\right]  }\right. \nonumber\\
&  \left.  -\left[  \left(  p_{lk}^{v^{\prime}c^{\prime}}\right)  ^{\ast
}p_{lk}^{v^{\prime}c^{\prime}}p_{ij}^{vc}\right]  ^{:\left[  -\mathbf{k}%
_{1}+\mathbf{k}_{2}+\mathbf{k}_{2}\right]  }\right. \nonumber\\
&  \left.  -\left[  \left(  p_{lk}^{v^{\prime}c^{\prime}}\right)  ^{\ast
}B_{lkij}^{v^{\prime}c^{\prime}vc}\right]  ^{:\left[  -\mathbf{k}%
_{1}+\mathbf{k}_{2}+\mathbf{k}_{2}\right]  }\right\} \nonumber\\
&  +\sum_{klv^{\prime}c^{\prime}}\left\{  \left(  \mathbf{\mu}_{il}%
^{vc^{\prime}}\right)  ^{\ast}\cdot\left[  \mathbf{E}\left(  t\right)  \left(
p_{kl}^{v^{\prime}c^{\prime}}\right)  ^{\ast}p_{kj}^{v^{\prime}c}\right]
^{\left[  -\mathbf{k}_{1}+\mathbf{k}_{2}+\mathbf{k}_{2}\right]  }\right.
\label{3rdO1}\\
&  \left.  +\left(  \mathbf{\mu}_{lj}^{v^{\prime}c}\right)  ^{\ast}%
\cdot\left[  \mathbf{E}\left(  t\right)  \left(  p_{lk}^{v^{\prime}c^{\prime}%
}\right)  ^{\ast}p_{ik}^{vc^{\prime}}\right]  ^{\left[  -\mathbf{k}%
_{1}+\mathbf{k}_{2}+\mathbf{k}_{2}\right]  }\right\}  ,\nonumber
\end{align}
To first order in the optical field, the single exciton density matrix $p$
contains either $\mathbf{k}_{1}$ or $\mathbf{k}_{2}$. \ Invoking the rotating
wave approximation (RWA) and by a perturbative expansion
\cite{Lindberg92,Weiser} to 3rd-order in the optical field, Eq. (\ref{3rdO1})
becomes
\begin{align}
&  -i\frac{\partial}{\partial t}p_{ij}^{vc:\left[  -\mathbf{k}_{1}%
+\mathbf{k}_{2}+\mathbf{k}_{2}\right]  }-\frac{i}{t_{ex}}p_{ij}^{vc:\left[
-\mathbf{k}_{1}+\mathbf{k}_{2}+\mathbf{k}_{2}\right]  }\nonumber\\
&  =-\sum_{n}T_{jn}^{c}p_{in}^{vc:\left[  -\mathbf{k}_{1}+\mathbf{k}%
_{2}+\mathbf{k}_{2}\right]  }\nonumber\\
&  -\sum_{m}T_{mi}^{v}p_{mj}^{vc:\left[  -\mathbf{k}_{1}+\mathbf{k}%
_{2}+\mathbf{k}_{2}\right]  }+V_{ij}p_{ij}^{vc:\left[  -\mathbf{k}%
_{1}+\mathbf{k}_{2}+\mathbf{k}_{2}\right]  }\nonumber\\
&  +\sum_{klv^{\prime}c^{\prime}}\left(  V_{kj}-V_{ki}-V_{lj}+V_{li}\right)
\nonumber\\
&  \cdot\left\{  \left[  \left(  p_{lk}^{v^{\prime}c^{\prime}:\mathbf{k}_{1}%
}\right)  ^{\ast}p_{lj}^{v^{\prime}c:\mathbf{k}_{2}}p_{ik}^{vc^{\prime
}:\mathbf{k}_{2}}\right]  \right. \nonumber\\
&  \left.  -\left[  \left(  p_{lk}^{v^{\prime}c^{\prime}:\mathbf{k}_{1}%
}\right)  ^{\ast}p_{lk}^{v^{\prime}c^{\prime}:\mathbf{k}_{2}}p_{ij}%
^{vc:\mathbf{k}_{2}}\right]  \right. \nonumber\\
&  \left.  -\left[  \left(  p_{lk}^{v^{\prime}c^{\prime}:\mathbf{k}_{1}%
}\right)  ^{\ast}B_{lkij}^{v^{\prime}c^{\prime}vc:\left[  \mathbf{k}%
_{2}+\mathbf{k}_{2}\right]  }\right]  \right\} \nonumber\\
&  -\mathbf{E}_{2}\left(  t\right)  \cdot\sum_{klv^{\prime}c^{\prime}}\left\{
\left(  \mathbf{\mu}_{il}^{vc^{\prime}}\right)  ^{\ast}\left(  p_{kl}%
^{v^{\prime}c^{\prime}:\mathbf{k}_{1}}\right)  ^{\ast}p_{kj}^{v^{\prime
}c:\mathbf{k}_{2}}\right. \nonumber\\
&  \left.  +\left(  \mathbf{\mu}_{lj}^{v^{\prime}c}\right)  ^{\ast}\left(
p_{lk}^{v^{\prime}c^{\prime}:\mathbf{k}_{1}}\right)  ^{\ast}p_{ik}%
^{vc^{\prime}:\mathbf{k}_{2}}\right\}  .\label{5}%
\end{align}

Apart from the density matrix element $p_{ij}^{vc:\left[  -\mathbf{k}%
_{1}+\mathbf{k}_{2}+\mathbf{k}_{2}\right]  }$, Eq. (\ref{5}) also contains the
matrix elements $p_{ij}^{vc\left[  \mathbf{k}_{1}\right]  }$, $p_{ij}%
^{vc\left[  \mathbf{k}_{2}\right]  }$ and $B_{lkij}^{v^{\prime}c^{\prime
}vc:\left[  \mathbf{k}_{2}+\mathbf{k}_{2}\right]  }$. \ Thus to solve for
$p_{ij}^{vc:\left[  -\mathbf{k}_{1}+\mathbf{k}_{2}+\mathbf{k}_{2}\right]
}(t_{3},t_{2},t_{1},t)$ at time $t$, we should calculate $p_{ij}^{vc\left[
\mathbf{k}_{1}\right]  }\left(  t\right)  $, $p_{ij}^{vc\left[  \mathbf{k}%
_{2}\right]  }\left(  t\right)  $ and $B_{lkij}^{v^{\prime}c^{\prime
}vc:\left[  \mathbf{k}_{2}+\mathbf{k}_{2}\right]  }\left(  t\right)  $
simultaneously. \ Following the similar procedure of deriving Eq. (\ref{5}),
we obtain the following equations for solving $p_{ij}^{vc\left[
\mathbf{k}_{1}\right]  }\left(  t\right)  $, $p_{ij}^{vc\left[  \mathbf{k}%
_{2}\right]  }\left(  t\right)  $ and $B_{lkij}^{v^{\prime}c^{\prime
}vc:\left[  \mathbf{k}_{2}+\mathbf{k}_{2}\right]  }\left(  t\right)  $: \ %

\begin{align}
&  -i\frac{\partial}{\partial t}p_{ij}^{vc\left[  \mathbf{k}_{1}\right]
}-\frac{i}{t_{ex}}p_{ij}^{vc\left[  \mathbf{k}_{1}\right]  }=-\sum_{n}%
T_{jn}^{c}p_{in}^{vc\left[  \mathbf{k}_{1}\right]  }-\sum_{m}T_{mi}^{v}%
p_{mj}^{vc\left[  \mathbf{k}_{1}\right]  }\nonumber\\
&  +V_{ij}p_{ij}^{vc\left[  \mathbf{k}_{1}\right]  }+\mathbf{E}_{1}\left(
t\right)  \cdot\left(  \mathbf{\mu}_{ij}^{vc}\right)  ^{\ast},\label{first}%
\end{align}%

\begin{align}
&  -i\frac{\partial}{\partial t}p_{ij}^{vc\left[  \mathbf{k}_{2}\right]
}-\frac{i}{t_{ex}}p_{ij}^{vc\left[  \mathbf{k}_{2}\right]  }=-\sum_{n}%
T_{jn}^{c}p_{in}^{vc\left[  \mathbf{k}_{2}\right]  }-\sum_{m}T_{mi}^{v}%
p_{mj}^{vc\left[  \mathbf{k}_{2}\right]  }\nonumber\\
&  +V_{ij}p_{ij}^{vc\left[  \mathbf{k}_{2}\right]  }+\mathbf{E}_{2}\left(
t\right)  \cdot\left(  \mathbf{\mu}_{ij}^{vc}\right)  ^{\ast},\label{second}%
\end{align}%
\begin{align}
&  -i\frac{\partial}{\partial t}B_{lkij}^{v^{\prime}c^{\prime}vc:\left[
\mathbf{k}_{2}+\mathbf{k}_{2}\right]  }-\frac{i}{t_{bi}}B_{lkij}^{v^{\prime
}c^{\prime}vc:\left[  \mathbf{k}_{2}+\mathbf{k}_{2}\right]  }\nonumber\\
&  =-\sum_{m}\left(  T_{jm}^{c}B_{lkim}^{v^{\prime}c^{\prime}vc:\left[
\mathbf{k}_{2}+\mathbf{k}_{2}\right]  }+T_{mi}^{v}B_{lkmj}^{v^{\prime
}c^{\prime}vc:\left[  \mathbf{k}_{2}+\mathbf{k}_{2}\right]  }\right.
\nonumber\\
&  \left.  +T_{km}^{c}B_{lmij}^{v^{\prime}c^{\prime}vc:\left[  \mathbf{k}%
_{2}+\mathbf{k}_{2}\right]  }+T_{ml}^{v}B_{mkij}^{v^{\prime}c^{\prime
}vc:\left[  \mathbf{k}_{2}+\mathbf{k}_{2}\right]  }\right) \nonumber\\
&  +\left(  V_{lk}+V_{lj}+V_{ik}+V_{ij}-V_{li}-V_{kj}\right)  B_{lkij}%
^{v^{\prime}c^{\prime}vc:\left[  \mathbf{k}_{2}+\mathbf{k}_{2}\right]
}\nonumber\\
&  -\left(  V_{lk}+V_{ij}-V_{li}-V_{kj}\right)  p_{ik}^{vc^{\prime}\left[
\mathbf{k}_{2}\right]  }p_{lj}^{v^{\prime}c\left[  \mathbf{k}_{2}\right]
}\nonumber\\
&  +\left(  V_{ik}+V_{lj}-V_{li}-V_{kj}\right)  p_{lk}^{v^{\prime}c^{\prime}
\left[  \mathbf{k}_{2}\right]  }p_{ij}^{vc\left[  \mathbf{k}_{2}\right]
}.\label{fourth}%
\end{align}

Solving the set of coupled equations (\ref{5}), (\ref{first})\ to
(\ref{fourth}) gives $p_{ij}^{vc:\left[  -\mathbf{k}_{1}+\mathbf{k}%
_{2}+\mathbf{k}_{2}\right]  }(t_{3},t_{2},t_{1},t)$. \ Substituting
$p_{ij}^{vc:\left[  -\mathbf{k}_{1}+\mathbf{k}_{2}+\mathbf{k}_{2}\right]
}(t_{3},t_{2},t_{1},t)$ into Eq. (\ref{2DD}) results in the $\mathbf{S}%
_{\mathbf{I}}$ 2D signal. \ Equations for the $\mathbf{S}_{\mathbf{II}}$ and
$\mathbf{S}_{\mathbf{III}}$ 2D signals can be derived in a similar way. \ 

\section{Numerical Calculation of Linear Absorption}

From Eqs. (\ref{interband})\ and (\ref{interpol}), the total polarization
along the $\mathbf{k}_{\mathbf{I}}$ direction is written as%

\begin{align}
\mathbf{P}\left(  t\right)   &  \mathbf{=}\mathbf{P}^{\left[  \mathbf{k}%
_{\mathbf{I}}\right]  }(t_{3},t_{2},t_{1},t)+c.c.\nonumber\\
&  =\sum_{ijvc}\mathbf{\mu}_{ij}^{vc}p_{ij}^{vc:\left[  \mathbf{k}%
_{\mathbf{I}}\right]  }(t_{3},t_{2},t_{1},t)+c.c.\nonumber\\
&  \equiv\mathbf{p}\left(  t\right)  e^{-i\omega_{g}t}+c.c.,
\end{align}
where $\mathbf{p}\left(  t\right)  $ is defined as the slowly-varying portion
relative to the bandgap frequency, $\omega_{g}$. \ Performing Fourier
transform of the polarization relative to $\omega_{g}$ gives%

\begin{align}
\mathbf{P}\left(  \omega\right)   &  =\int_{-\infty}^{\infty}\mathbf{P}\left(
t\right)  e^{-i\left(  \omega-\omega_{g}\right)  t}dt\nonumber\\
&  =\int_{-\infty}^{\infty}\left[  \mathbf{p}\left(  t\right)  e^{-i\omega
_{g}t}+c.c.\right]  e^{-i\left(  \omega-\omega_{g}\right)  t}dt\nonumber\\
&  =\int_{-\infty}^{\infty}\left[  \mathbf{p}\left(  t\right)  e^{-i\omega
t}+\mathbf{p}^{\ast}\left(  t\right)  e^{-i\left(  \omega-2\omega_{g}\right)
t}\right]  dt\nonumber\\
&  \approx\int_{-\infty}^{\infty}\mathbf{p}\left(  t\right)  e^{-i\omega
t}dt.\label{pw}%
\end{align}

Similarly, performing Fourier transform of the $j-$th optical pulse (Eq.
(\ref{efield})) relative to the bandgap frequency $\omega_{g}$ gives (spatial
portion is omitted)
\begin{align}
\mathbf{E}_{\mathbf{opt}}(\omega)  &  =\int_{-\infty}^{\infty}\left[
\mathcal{E}_{j}^{+}e^{-i\omega_{j}t}+c.c.\right]  e^{-i\left(  \omega
-\omega_{g}\right)  t}\nonumber\\
&  =\int_{-\infty}^{\infty}\left[  \mathcal{E}_{j}^{+}e^{-i\left(  \omega
_{j}-\omega_{g}\right)  t}e^{-i\omega_{g}t}e^{-i\left(  \omega-\omega
_{g}\right)  t}+e^{-i\left(  \omega-\omega_{g}\right)  t}\cdot c.c.\right]
dt\nonumber\\
&  \approx\int_{-\infty}^{\infty}\mathcal{E}_{j}^{+}e^{-i\left(  \omega
_{j}-\omega_{g}\right)  t}e^{-i\omega t}dt.\label{Ew}%
\end{align}
where $\mathbf{E}_{\mathbf{opt}}(\omega)$ is approximately the Fourier
transform of pulse envelope $\mathcal{E}_{j}^{+}$ because we consider only
resonant excitation where $\omega_{j}$ is close to $\omega_{g}$. \ Finally,
the linear absorption is calculated as\cite{298,Lijun06}%
\begin{equation}
\alpha\left(  \omega\right)  =\frac{\omega_{j}}{n^{\prime}c\epsilon_{0}%
}\operatorname{Im}\left[  \frac{\mathbf{P}\left(  \omega\right)
\cdot\mathbf{E}_{\mathbf{opt}}^{\ast}(\omega)}{\left|  \mathbf{E}%
_{\mathbf{opt}}(\omega)\right|  ^{2}}\right]  ,\label{absorption}%
\end{equation}
where $n^{\prime}$ is the average, frequency-independent refractive index of
the quantum well, $\epsilon_{0}$ is the vacuum permittivity and $c$ is the
speed of light. \

Figure Captions

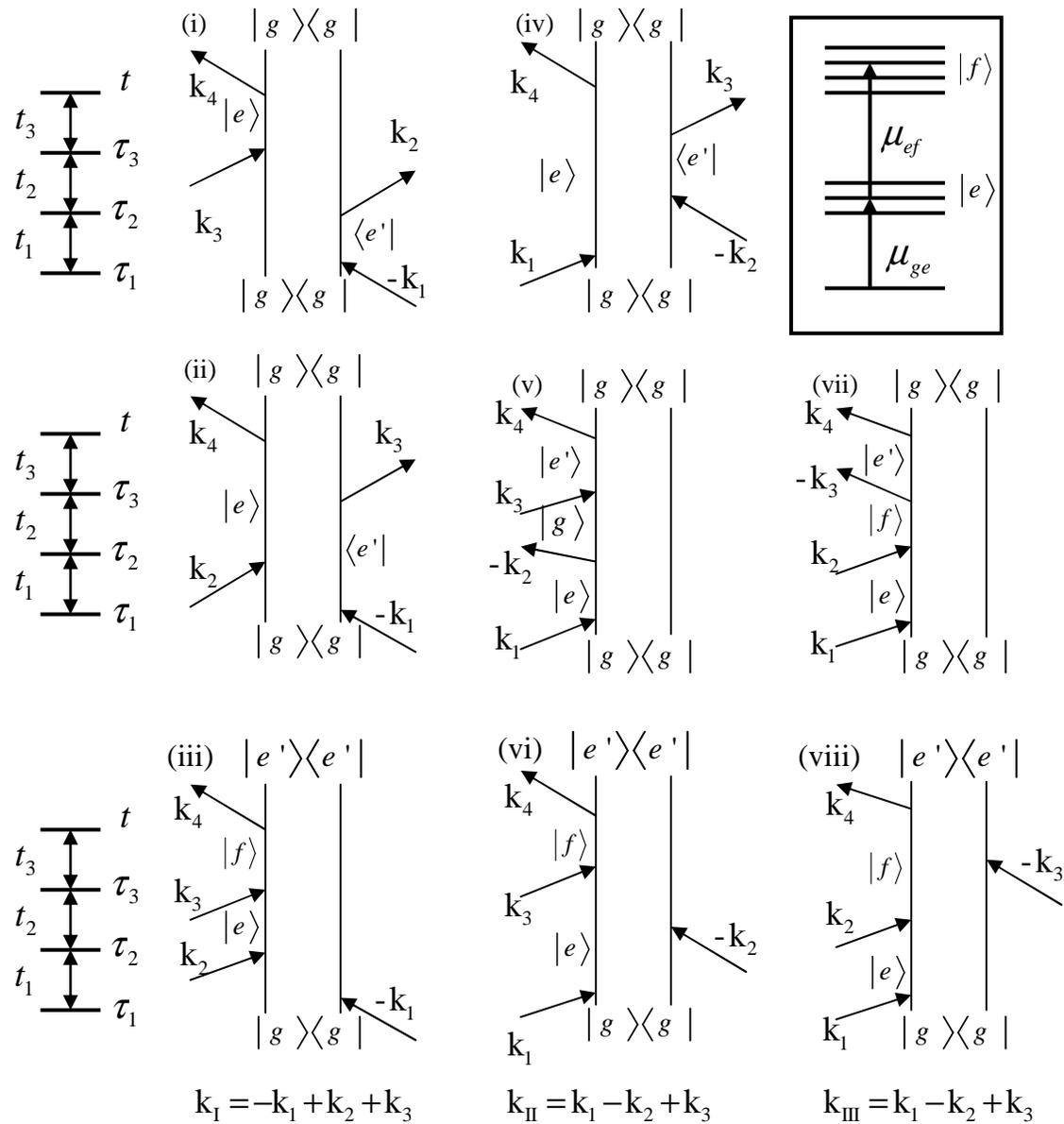
Fig. 1. Feynman diagrams for the possible coherent 2D spectroscopic techniques
for the exciton level scheme given in the top right. \ $|g\rangle$ is the
ground state, $|e\rangle$ is the single exciton manifold and\ $|f\rangle$ is
the two-exciton manifold. \ $\mu_{ge}$ and $\mu_{ef}$ are the corresponding
transition dipoles. [From Ref. \cite{paper386}] \ 

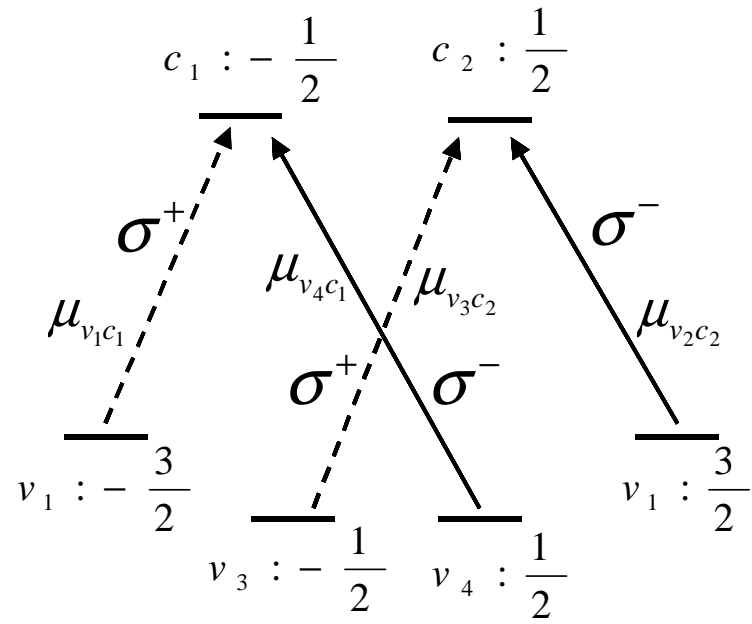
Fig. 2. \ Selection rules for III-V semiconductors. $\ c_{1}$ and $c_{2}$ are
two conduction bands whose spins are $s_{z}=-\frac{1}{2}$ and $s_{z}=\frac
{1}{2}$ respectively. \ $v_{1}$ and $v_{2}$ are HH valence bands with spins
$s_{z}=-\frac{3}{2}$ and $s_{z}=\frac{3}{2}$ respectively. \ $v_{3}$ and
$v_{4}$ are LH valence bands with spins $s_{z}=-\frac{1}{2}$ and $s_{z}%
=\frac{1}{2}$ respectively. \ 

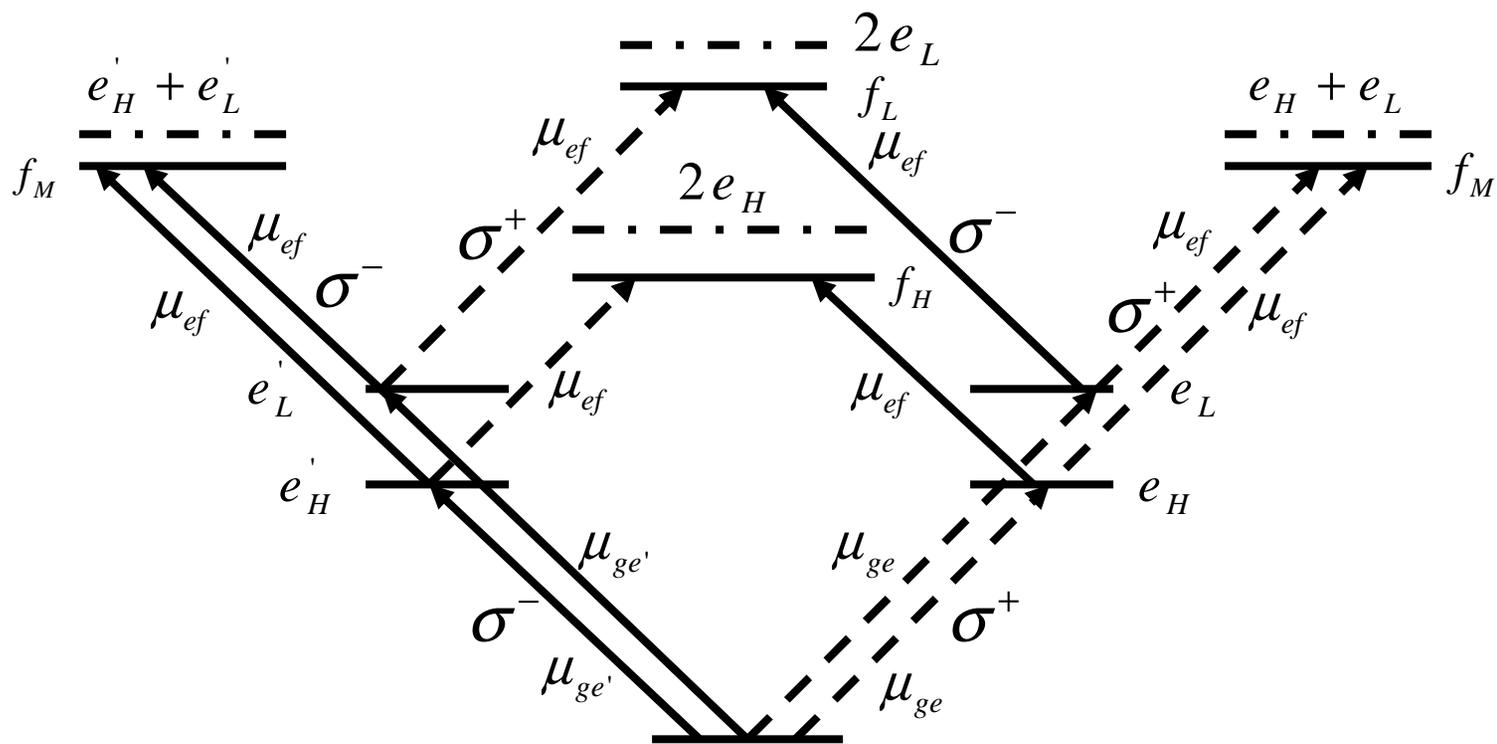
Fig. 3. \ Single-exciton and bound two-exciton optical transitions in III-V semiconductors.

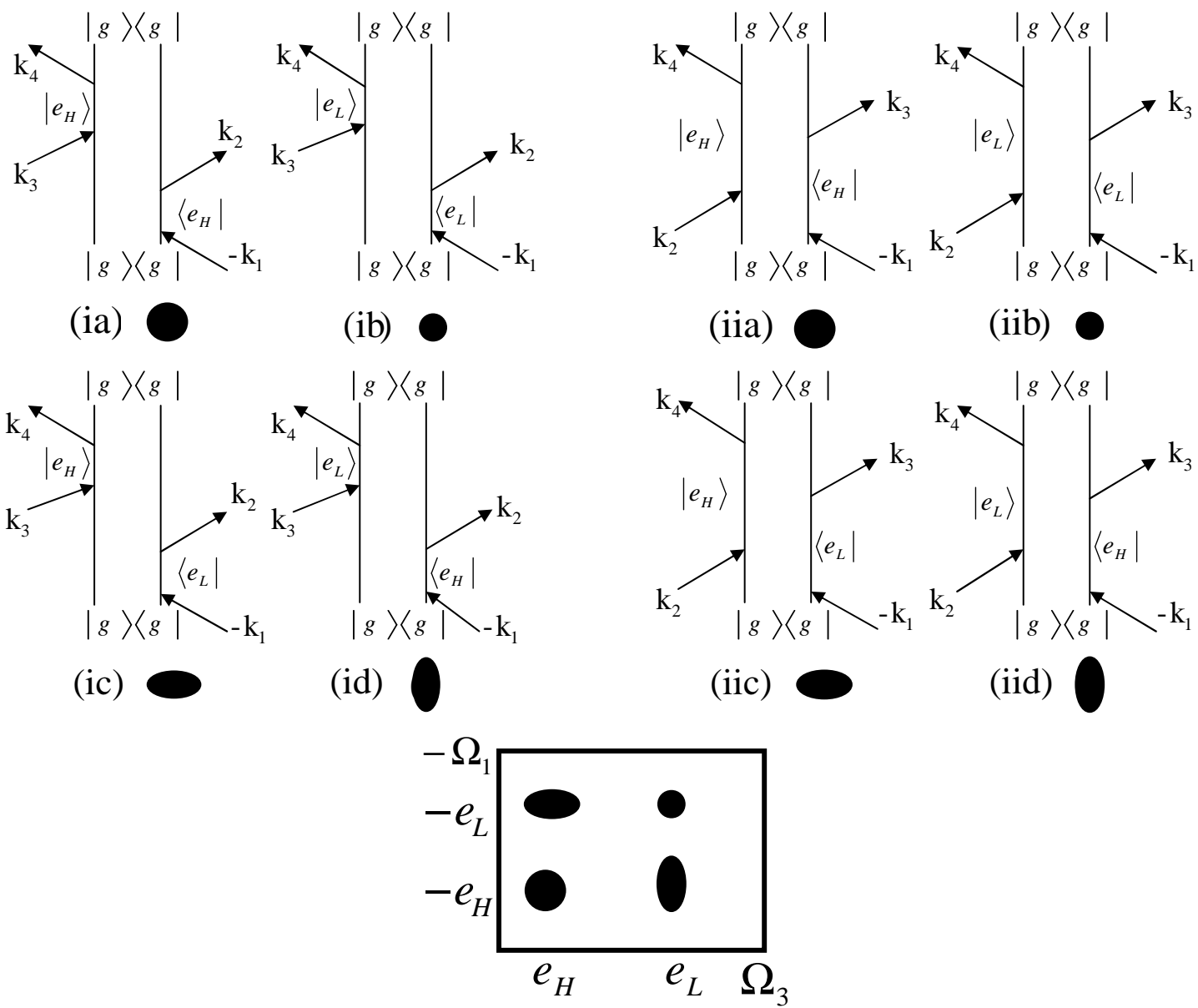
Fig. 4. \ Top:\ the Feynman diagrams derived from Fig. 1-(i) and (ii).
\ Bottom: the schematic 2D spectrum.

Fig. 5. Top:\ the Feynman diagrams derived from Fig. 1-(iii). \ Bottom: the
schematic 2D spectrum.

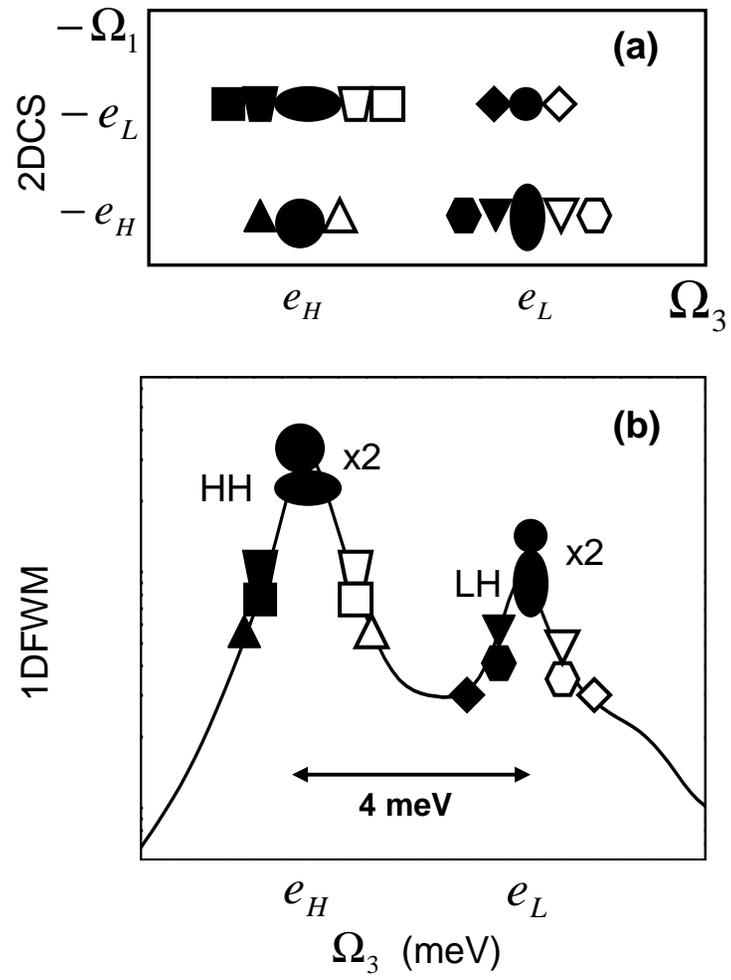
Fig. 6. (a)\ The total schematic 2D spectrum of $\mathbf{k}_{\mathbf{I}%
}=-\mathbf{k}_{1}+\mathbf{k}_{2}+\mathbf{k}_{3}$ (diagrams (i), (ii) and
(iii)). \ (b) The 1DFWM signal $\left|  \mathbf{S}_{\mathbf{I}}\left(
\Omega_{3},t_{2},t_{1}\right)  \right|  $ displayed versus $\Omega_{3}$\ for
fixed $t_{2}$ and $t_{1}$. \ 

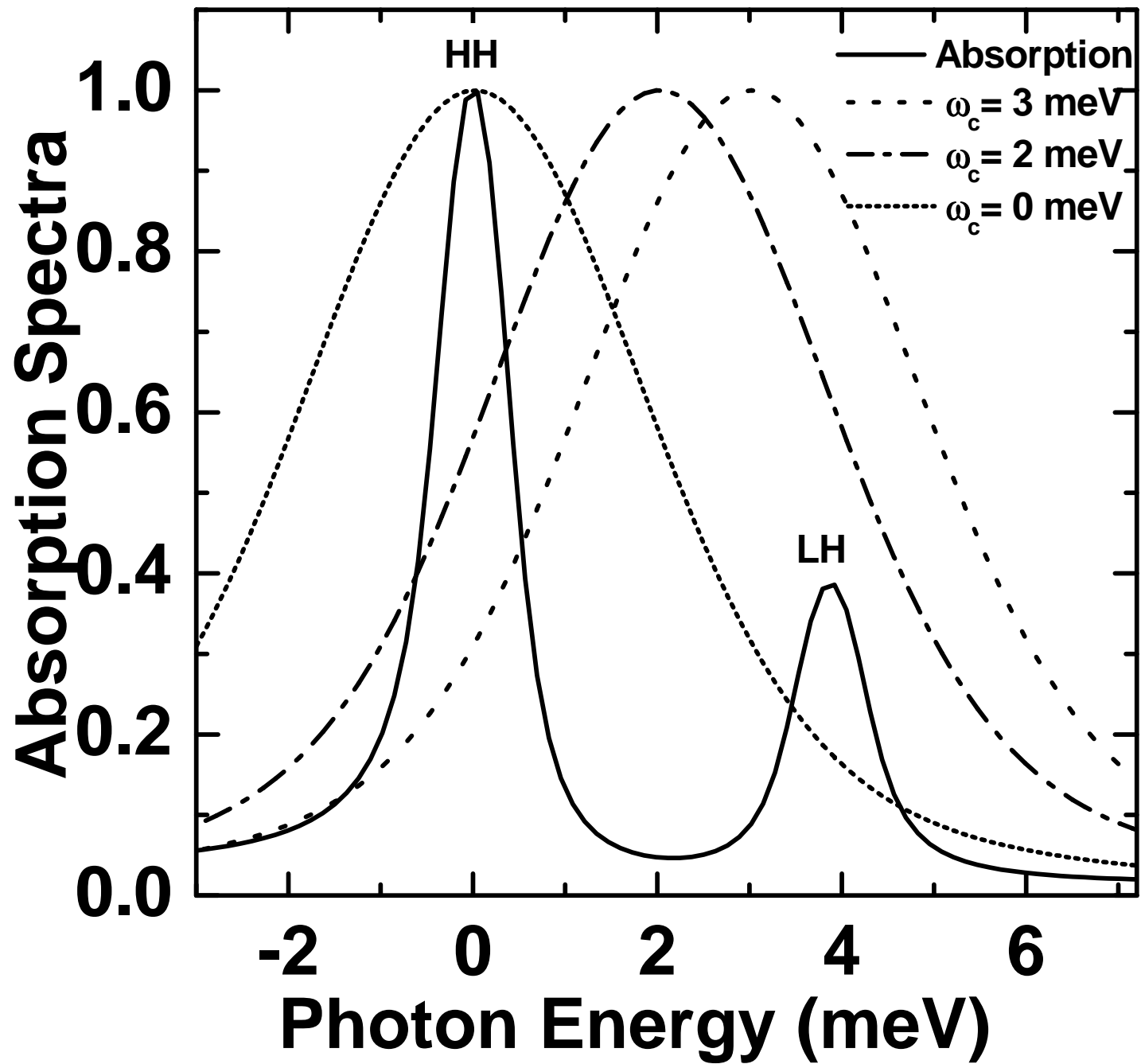
Fig. 7. \ (solid): Simulated linear absorption of the single quantum well (see
text for parameters). \ Also shown are optical pulse power spectra $\left|
\mathbf{E}_{\mathbf{opt}}(\omega)\right|  ^{2}$ (Eqs. (\ref{Ew})) for
$\omega_{c}=3$ meV(dotted), $\omega_{c}=2$ meV(dash dot) and $\omega_{c}=0$
meV (short dot). \ 

Fig. 8. \ Convergence of the modulus $\left|  \mathbf{S}_{\mathbf{I}}%
(\Omega_{3},t_{2},t_{1})\right|  $ with basis size. \ The 0-6 meV region has
converged for $N=10$. The higher energy continuum has not converged. \ 

Fig. 9. (Color online)$\ $(A) The modulus $\left|  \mathbf{S}_{\mathbf{I}%
}(\Omega_{3},t_{2},-\Omega_{1})\right|  $ for co-linearly polarized pulses
(XX) ($\omega_{c}=3$) meV; \ (B) same as (A) but excluding two-exciton
contributions; (C) same as (A) but with $\sigma^{+}\sigma^{+}$ excitation. \ 

Fig. 10. (Color online) (A) The modulus $\left|  \mathbf{S}_{\mathbf{I}%
}(\Omega_{3},t_{2},-\Omega_{1})\right|  $ calculated with co-linearly
polarized pulse excitations ($\omega_{c}=2$ meV); \ (B) same as (A) but
excluding the two-exciton contributions. \ 

Fig. 11. (Color online)\ (A) The modulus $\left|  \mathbf{S}_{\mathbf{I}%
}(\Omega_{3},t_{2},-\Omega_{1})\right|  $ calculated with cross-linearly (XY)
polarized pulse excitations ($\omega_{c}=0$ meV); \ (B) same as (A) but
excluding the two-exciton contributions. \ 

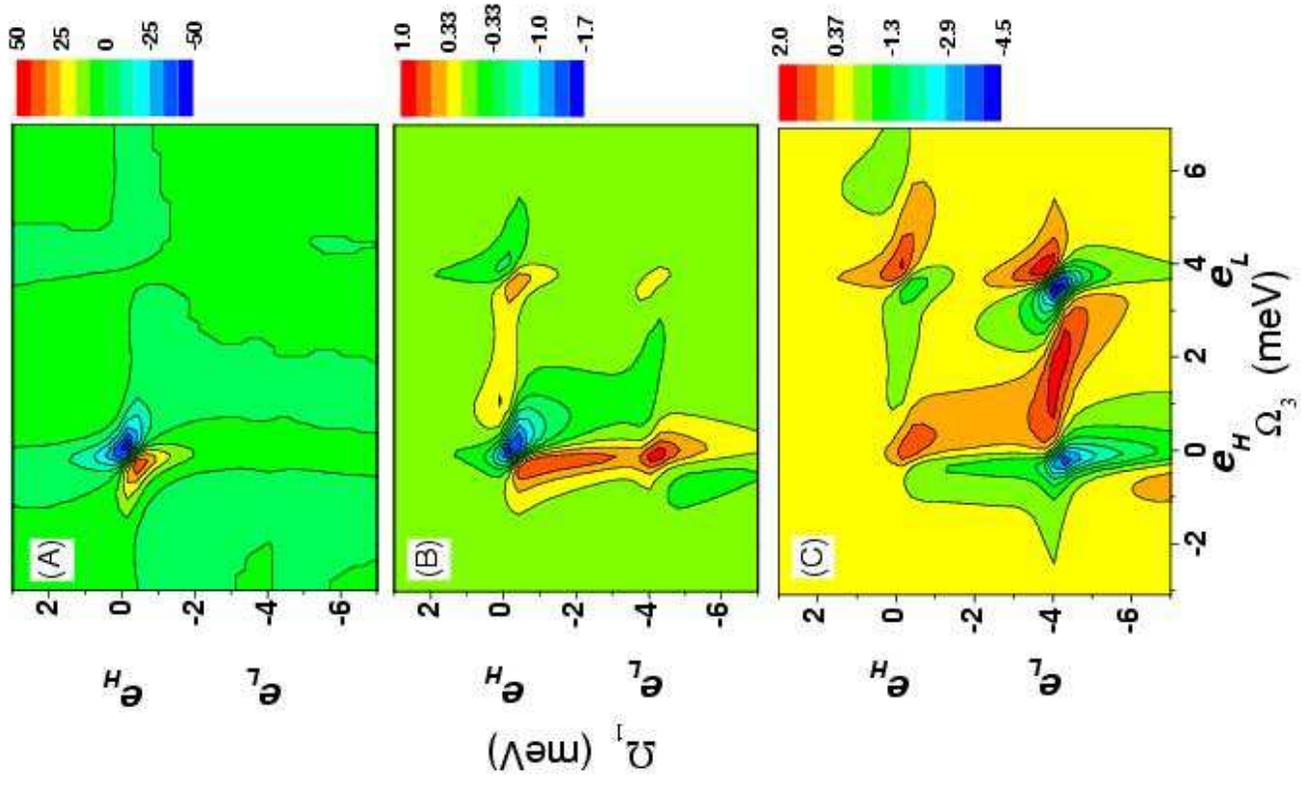
Fig. 12. (Color online) Re($\mathbf{S}_{\mathbf{I}}(\Omega_{3},t_{2}%
,-\Omega_{1}))$ calculated with XX excitation for (A)\ $\omega_{c}=0$ meV;
\ (B) $\omega_{c}=2$ meV; and (C) $\omega_{c}=3$ meV.\ 

Fig. 13. \ The $\mathbf{S}_{\mathbf{I}}\left(  \Omega_{3},\Omega_{2}%
,t_{1}\right)  $ signal. \ (a) Diagrams (ia), (ib), (ic) and (id). \ (b)
Diagrams (iia), (iib) and (iic). \ (c) Diagrams (iiia) to (iiid). \ (d) Total
spectrum. \ 
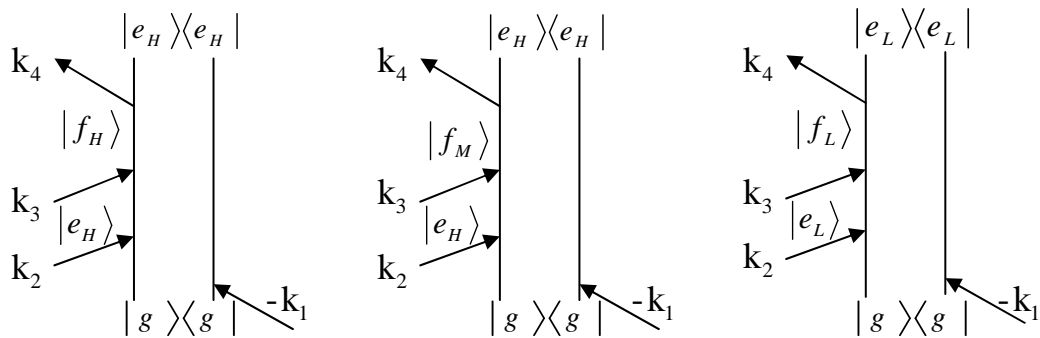
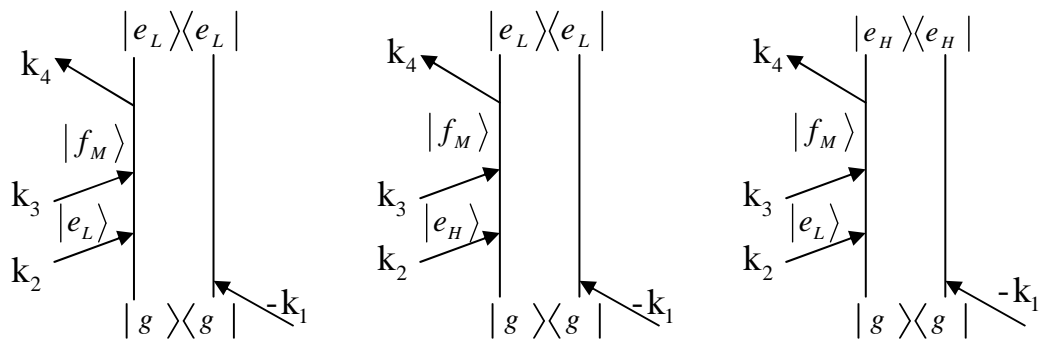
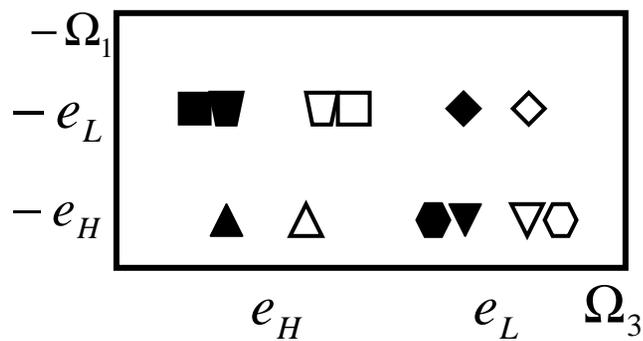
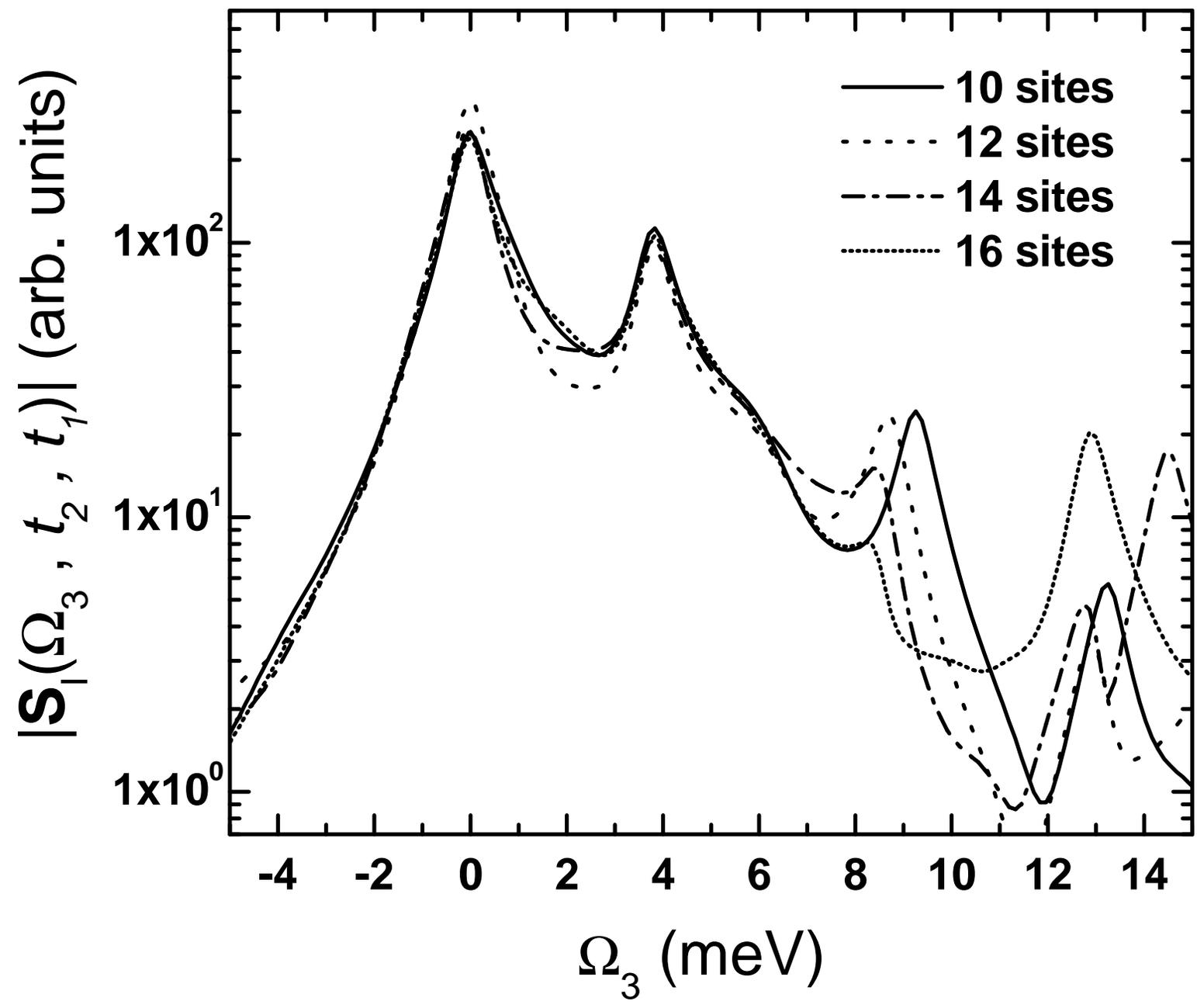
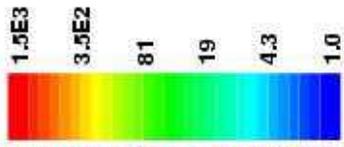
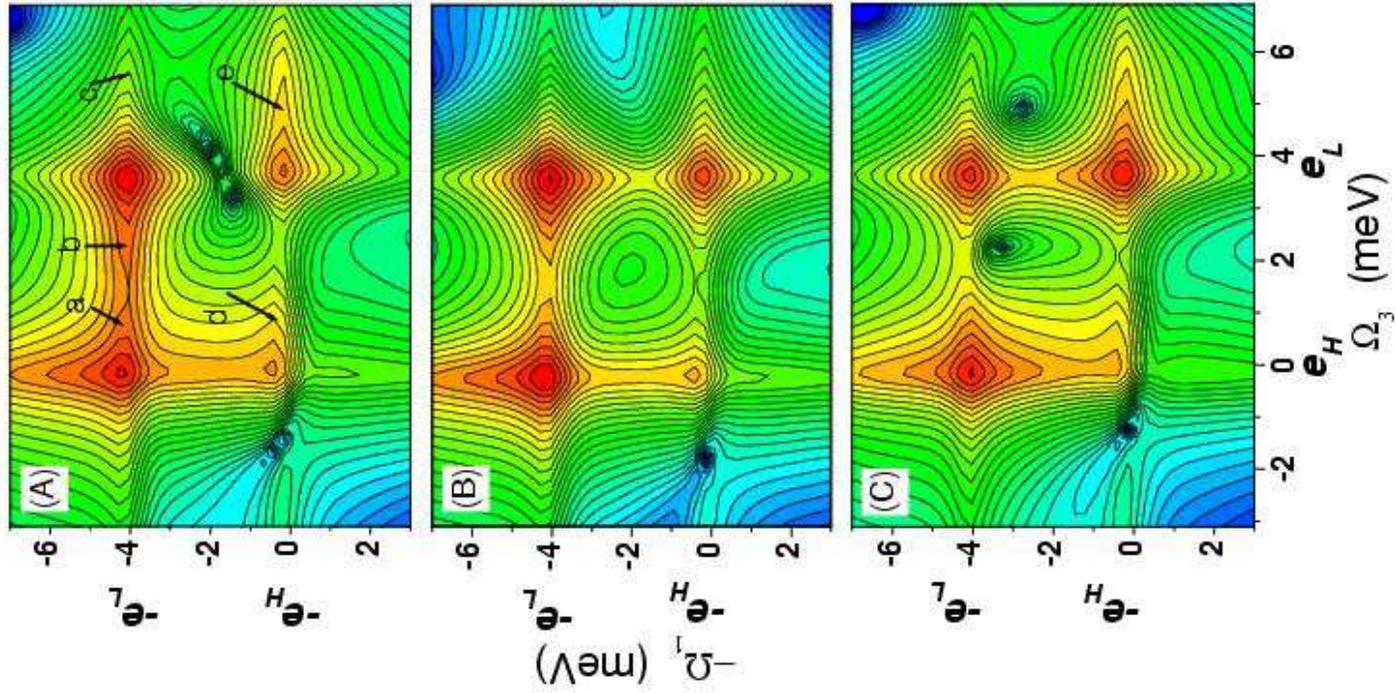
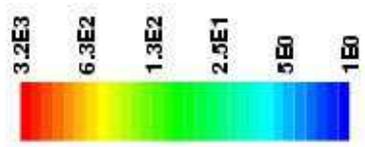
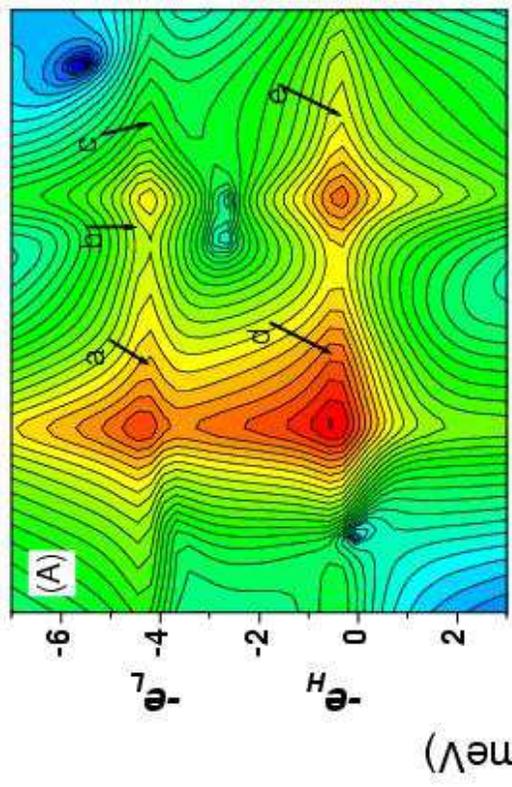
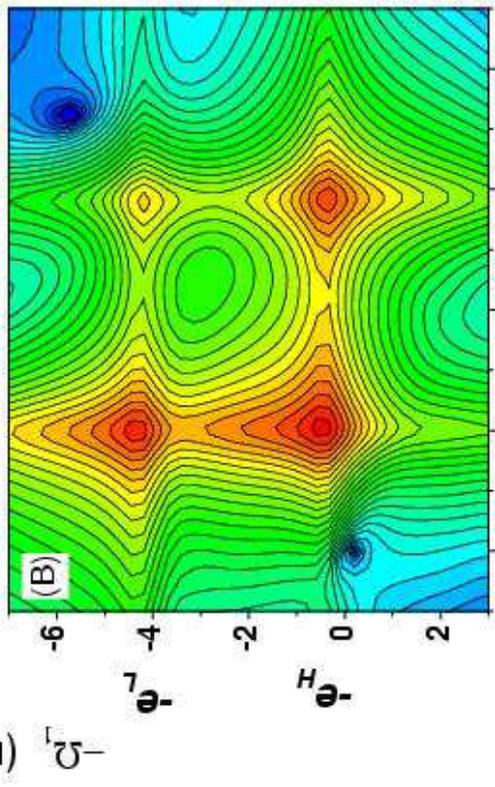
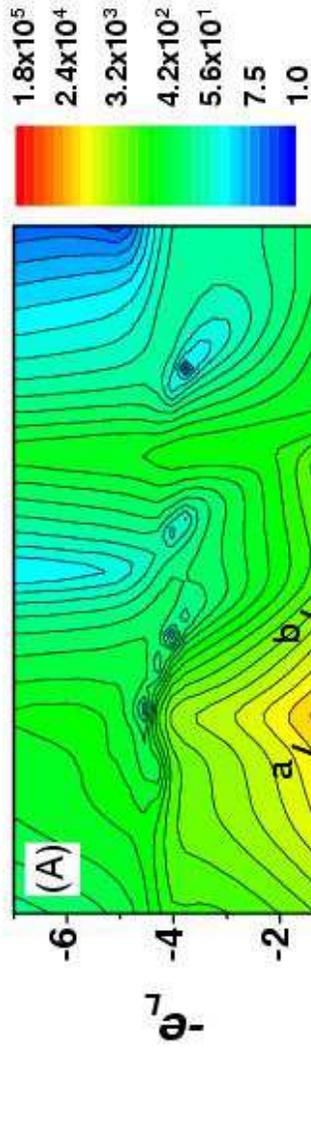
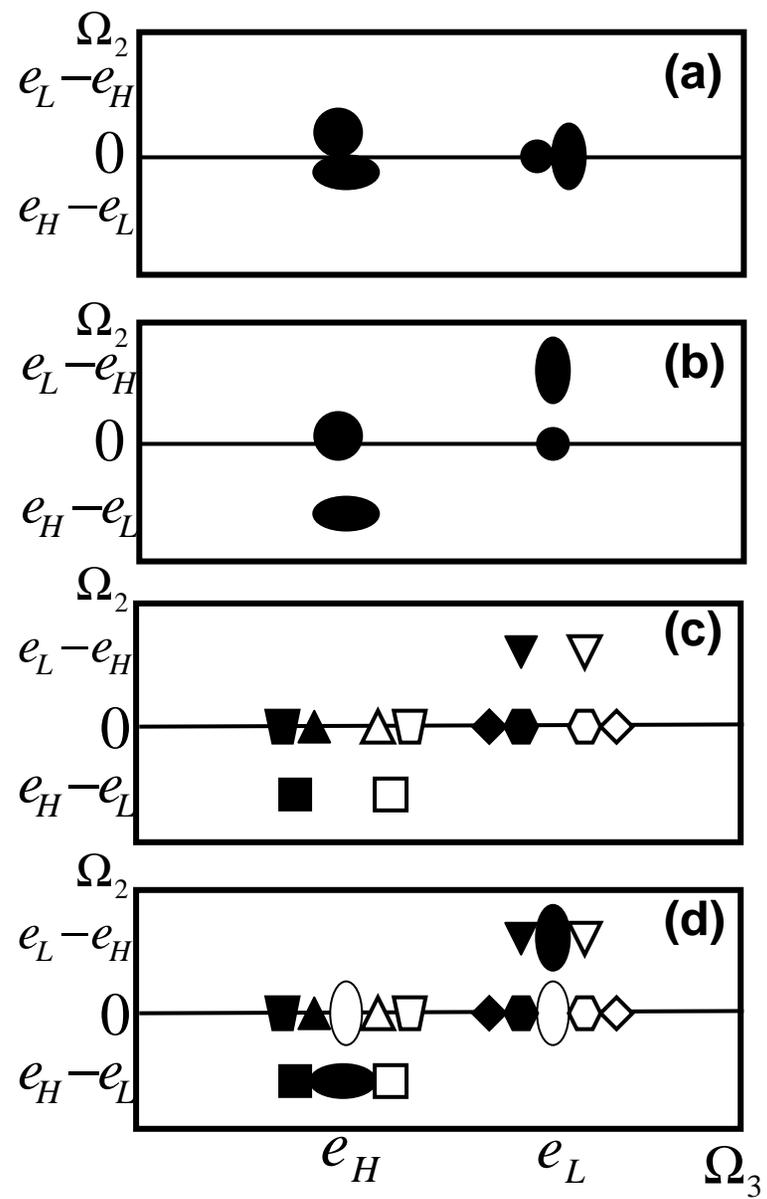
\end{document}